\documentclass[preprint,12pt]{aastex}

\begin{document}

\title{Quasi-local evolution of cosmic gravitational clustering in
       weakly non-linear regime}

\author{Jes\'us Pando,\altaffilmark{1}
\ Longlong Feng\altaffilmark{2}
\ and 
Li-Zhi Fang\altaffilmark{3}
}

\altaffiltext{1}{Department of Chemistry and Physics, Chicago State
   University, Chicago, IL 60628}
\altaffiltext{2}{Center for Astrophysics, University of Science and
  Technology of China, Hefei, Anhui 230026,
  National Astronomical Observatories, Chinese Academy of Science, 
  Chao-Yang District, Beijing, 100012, P.R. China}
\altaffiltext{3}{Department of Physics, University of Arizona, Tucson,
  AZ 85721}

\begin{abstract}

We investigate the weakly non-linear evolution of cosmic gravitational 
clustering in phase space by looking at the Zel'dovich solution in the
discrete wavelet transform (DWT) 
representation. We show that if the initial perturbations are Gaussian, 
the relation between the evolved DWT mode and the initial perturbations 
in the weakly non-linear regime is quasi-local. That is, the evolved 
density perturbations are mainly determined by the initial 
perturbations localized in the same spatial range. Furthermore, we 
show that the evolved mode is monotonically related to the initial 
perturbed mode. Thus large (small) perturbed modes statistically 
correspond to the large (small) initial perturbed modes. We test this 
prediction by using QSO Ly$\alpha$ absorption 
samples. The results show that the  weakly non-linear features for 
both the transmitted flux and identified forest lines are 
quasi-localized. The locality and monotonic properties provide a solid 
basis for a DWT scale-by-scale Gaussianization reconstruction 
algorithm proposed by Feng \& Fang (Feng \& Fang, 2000) for data in 
the weakly non-linear regime. With the Zel'dovich solution, we find 
also that the major non-Gaussianity caused by the weakly non-linear 
evolution is local scale-scale correlations. Therefore, to have a precise 
recovery of the initial Gaussian mass field, it is essential to remove 
the scale-scale correlations. 

\end{abstract}


\section{Introduction}

One of the basic goals of large scale structure study is to reconstruct
the initial mass field of the universe. Assuming observed objects trace
the underlying matter field in some way, it should be possible to 
recover the initial conditions of the mass field. For instance, if the 
probability distribution function (PDF) of the initial mass 
fluctuations is Gaussian, one may recover the initial Gaussian mass 
fluctuations by properly removing all non-Gaussian features via the 
Gaussianization reconstruction algorithm (Weinberg 1992.)

The key step in the Gaussianization algorithm is a mapping from a 
smoothed observed non-Gaussian distribution of the density
contrast, $\delta(x)= (\rho -\bar{ \rho})/ \bar{\rho}$ ($\rho$ is
the mass density, and $\bar{\rho}$ is its mean) into a smoothed initial 
Gaussian density contrast $\delta_0(x)$. The basic assumptions of this 
mapping are that the relation between $\delta(x)$ and the initial density 
distribution $\delta_0(x)$ is local and monotonic, i.e., the rank order of 
the mass density field, smoothed over a given scale, is preserved even under 
nonlinear gravitation evolution(Narayanan \& Weinberg 1998). Thus, the 
high initial density pixels will evolve into high $\delta(x)$ pixels and 
low initial density pixels into low $\delta(x)$ pixels. With this 
assumption, the Gaussian mapping can be realized by a point-to-point, 
order-preserving transformation. The shape of the initial Gaussian 
density field is recovered with an arbitrary normalization.

This method has been employed to recover the initial mass field and 
power spectrum from galaxy redshift surveys and from the transmitted 
flux of the Ly$\alpha$ absorption in QSO spectra (Croft et al. 1998.) 
The validity of the point-to-point (or pixel-to-pixel) recovery 
of the initial Gaussian mass field from the evolved mass field stems 
from the belief that the transmitted flux probably is a pixel-to-pixel 
tracer of the underlying dark matter distribution (Bi 1993; Fang et al. 
1993; Bi, Ge \& Fang 1995; Hernquist et al 1996; Bi \& Davidsen 1997; 
Hui, Gnedin \& Zhang 1997.)

However, whether the order-preserving assumption is reasonable is far 
from clear. It has been argued that the order-preserving condition may 
be a poor approximation to the actual dynamics because of the 
non-locality of gravitational evolution (Monaco \& Efstathiou 2000). 
Gravitational clustering is not localized. Even in the weakly 
non-linear evolutionary regime, the processes of cosmic clustering 
typically are those of objects free falling into potential wells 
(e.g. Xu, Fang \& Wu 2000), Fourier mode-mode coupling 
(e. g. Suto \& Sasaki 1991), and the merging of pre-virialized dark 
halos. These processes are generally {\it non-local}.

In the Zel'dovich approximation (Zel'dovich 1970), the density field 
$\rho({\bf x}, t)$ at (Eulerian) comoving position ${\bf x}$ and time 
$t$ is determined by the initial perturbation at (Lagrangian) comoving 
position, ${\bf q}$, plus a displacement ${\bf S}$:
\begin{equation}
{\bf x}({\bf q}, t)= {\bf q} + {\bf S}({\bf q}, t).
\end{equation}
The displacement ${\bf S}({\bf q}, t)$ represents the effect of density
perturbations on the trajectories of gravitating particles. Therefore, 
cosmic self-gravitating systems do not follow a Eulerian 
{\it point-to-point} localized evolution. Even when the transmitted 
flux is locally determined by the evolved underlying dark matter 
distribution, the relation between the initial mass field and the 
transmitted flux might still be non-local. The initial power spectrum 
reconstructed from the point-to-point Gaussian mapping of Ly$\alpha$ 
transmitted flux shows a systematic suppression on small scales
(Croft et al. 1999.) This problem may be mitigated by smoothing the 
data. However, the smoothing scale is put in by hand and it is hard 
to understand how this scale can be determined from the dynamics of 
the non-local evolution.  For these reasons the basis of the 
order-preserving transformation and the point-to-point Gaussian 
mapping need to be reconsidered.

In this paper, we will show first that the assumption of point-to-point
locality is not necessary for a proper Gaussianization reconstruction.

Recently, we showed that the transmitted flux Gaussianized by
the point-to-point Gaussian mapping is still largely non-Gaussian
(Feng \& Fang 2000.) That is, the non-Gaussianities of the original
transmitted flux are retained in the mass field recovered by the
point-to-point, order-preserving transformation. Especially troubling
is the fact that the scale-scale correlations of the Gaussianized 
field are about the same as before the mapping.
The mass power spectrum recovered by the point-to-point Gaussian 
mapping is systematically lower than the initial mass spectrum on 
scales at which the scale-scale correlation of the recovered mass 
field are substantial.

To solve this problem, a scale-by-scale algorithm of the Gaussian 
mapping was proposed (Feng \& Fang 2000.) In this algorithm the 
order-preserving transformation is done on the coefficients of a 
space-scale decomposition rather than on each pixel of the transmitted 
flux. This method can effectively remove the non-Gaussianities of the 
observed distribution, especially clearing the scale-scale 
correlations. Consequently, small scale
suppression is significantly reduced. Furthermore, the scale-by-scale 
Gaussianization procedure does not require a point-to-point order 
preservation, but only the order preservation of the density 
fluctuations of the space-scale decomposed field. 

We will show that the order preservation 
assumption for the scale-by-scale Gaussian mapping is supported by 
the Zel'dovich approximation and that in quasi-linear evolution, 
the gravitational clustering is quasi-local and monotonic. Moreover, 
we will find that the non-Gaussian features during the weakly 
non-linear regime are dominated by the local scale-scale correlations. 
Therefore, at least for the transmitted flux and forest lines of the 
QSO Ly$\alpha$ absorption spectrum, the scale-by-scale Gaussianization 
algorithm has a solid basis in the weakly non-linear regime.

The paper is organized as follows: Section 2 presents the quasi-locality 
condition needed for the scale-by-scale Gaussianization recovery. In 
Section 3 we show the quasi-locality and monotonic nature of the weak 
non-linear evolution using the Zel'dovich solution of the growth modes. 
Section 4 presents the result of measuring the quasi-locality for 
samples of the Ly$\alpha$ absorption spectrum. The conclusions are 
summarized in Section 5.

\section{Scale-by-scale Gaussian mapping and the quasi-locality 
condition}

\subsection{Problems with the point-to-point Gaussian mapping}

Let us briefly introduce the algorithm for a point-to-point Gaussian 
mapping which is designed for recovering the initial Gaussian mass 
density contrast $\delta_0({\bf x})$ from an observed, non-Gaussian 
distribution $\delta({\bf x})$. As an example, we consider an 
observed transmitted flux $F$ or absorption optical depth 
$\tau = - \ln F$ of the Ly$\alpha$ 
absorption in QSO spectra. The distribution is one--dimensional and of 
length $L$. The PDF of the observed transmitted flux $F$ is generally 
non-Gaussian, while the PDF of the initial density contrast $\delta_o$ is 
assumed to be Gaussian in a large variety of structure formation models. 
The relation between $F=e^{-\tau}$ and $\delta_0$ is assumed to be
order-preserved, or monotonic, i.e., high initial density 
$\delta_0$ pixels evolve into high absorption optical depth $\tau$ 
pixels, while low initial density pixels into low $\tau$ pixels. Thus, 
a point-to-point (or pixel-to-pixel) Gaussian mapping can be done as 
follows.  First, sort in ascending order, the $N$ pixels by the 
observed amount of flux. Second, assign to the n-th pixel a density 
contrast $\delta$ by the solving the error function equation 
$(2\pi)^{-1/2}\int_{-\infty}^{\delta}\exp(-x^2/2)dx= n/N$. This 
pixel-to-pixel Gaussian mapping produces a mass field with the same rank
order as the original flux but with a Gaussian PDF. After determining 
the overall amplitude of the Gaussian mapped field by a separate 
procedure, the initial density field $\delta_0$ is recovered 
(Weinberg 1992.)

However, the density field recovered by the pixel-to-pixel Gaussian 
mapping still exhibits non-Gaussian features. Especially, the 
scale-scale correlations of the Gaussianized field are almost as 
strong as the pre-Gaussianized flux. The pixel-to-pixel Gaussian 
mapping does not remove all the non-Gaussianities and the initial 
Gaussian field will not be recovered precisely. The power spectrum 
recovered by the point-to-point Gaussian mapping is systematically 
lower than the initial mass spectrum on scales at which the scale-scale 
correlation is substantial. Physically, the effect might just be due to
the fact that $\delta$ is simply too non-linear, and the relation 
between $\delta$ and $\delta_0$ is no longer monotonic. If this is the case,
then a scale-adaptive smoothing is required before Gaussianization.

However, a smoothing window function sometimes introduces spurious features.
For instance, the aliasing of the power spectrum on scales around the 
Nyquist frequency corresponds to the scale of the smoothing grid.
However with the DWT analysis one need not introduce window functions 
for smoothing because the scaling functions of the wavelet transform 
already play that role. The scaling function is orthogonal to the 
relevant wavelets. False correlations due to the smoothing window can 
then be perfectly eliminated (Feng \& Fang 2000.) Therefore,
a scale-adaptive smoothing and reconstruction can be realized by 
the scale-by-scale Gaussian mapping.

\subsection{Scale-by-scale Gaussian mapping}

Scale-by-scale, or scale-adaptive Gaussianization can be performed 
via the discrete wavelet transform (DWT)\footnote{For the details of the  
DWT refer to the classic papers by Mallat (1989a,b); 
Meyer (1992); Daubechies, (1992) and references therein, and for 
physical applications, refer to Fang \& Thews (1998) and references
therein.}. The first step in this process is to wavelet 
expand the flux $F$ as
\begin{equation}
F(x) = \bar{F} + \sum_{j=0}^{\infty} \sum_{l= 0}^{2^j -1}
\tilde{\epsilon}_{j,l} \psi_{j,l}(x)
\end{equation}
where $\psi_{j,l}(x)$, $j=0,1,...$, $l=0...2^j-1$, is an orthogonal
and complete basis. The wavelet basis (mode) $\psi_{j,l}(x)$ is 
localized both in physical space and in Fourier (scale) space. The 
function $\psi_{j,l}(x)$ is centered at position $lL/2^j$ in physical 
space and at wavenumber $2\pi\times 2^j/L$ in Fourier space. The wavelet
function coefficients (WFCs), $\tilde{\epsilon}_{j,l}$, are labeled by
the two subscripts $j$ and $l$ corresponding to the scale and position
respectively. The $\tilde{\epsilon}_{j,l}$ describe the fluctuation of 
the  flux on scale $L/2^j$ at position $lL/2^j$ and are computed as the
inner product 
\begin{equation}
\tilde{\epsilon}_{j,l}= \int_0^L F \psi_{j,l}dx.
\end{equation}

The density fields in cosmology are assumed to be ergodic, that is, the 
average over an ensemble is equal to the spatial average taken over {\it one}
realization. This is the so-called fair sample hypothesis (Peebles 1980).
A homogeneous Gaussian field with a continuous power spectrum is certainly 
ergodic(Adler 1981). In some non-Gaussian cases, such as homogeneous and 
isotropic turbulence, ergodicity also approximately holds(Vanmarke, 1983). 
Roughly, the ergodic assumption is reasonable if the spatial correlations 
decrease sufficiently rapidly with increasing separation. This 
property can effectively be used by the DWT because the wavelet
$\psi_{j,l}(x)$ are orthogonal with respect to the position index $l$. 
The SFCs $\tilde{\epsilon}_{j,l}$ ($j=0...2^j-1$) can be considered  
as $2^j$ independent sampling, without false correlations caused by 
a non-orthogonal or redundant decomposition. Thus, for ergodic random fields 
the $2^j$ WFCs form a statistical ensemble on scale $j$. So for quantity 
$X_{j}$, the ensemble average is given by  
\begin{equation}
\langle X_j \rangle \simeq 
 \frac{1}{2^j}\sum_{l=0}^{2^j-1} X_{j,l}
\end{equation}
where $X_{j,l}$ is the quantity at position $l$. In other words, 
the distribution of the $2^j$ WFCs is the one-point distribution
of $X_j$.  

In short we have arrived at the scale-by-scale Gaussianization algorithm 
which consists of the following steps: 1.) Sort out the $2^j$ WFCs, 
$\tilde{\epsilon}_{j,l}$, of the flux field $F$ in ascending order. 
2.) Assign to the $2^j$ WFCs, the Gaussianized density contrast 
$\delta^g$ at position $l$ by the solving the error function equation. 
In this algorithm, the order-preserving transformation is performed 
scale-by-scale. It gives a position($l$)-to-position($l$) Gaussian 
mapping for each scale $j$. After a proper amplitude normalization 
(Feng \& Fang 2000), the initial Gaussian density field $\delta_0$ is 
reconstructed with the Gaussianized WFCs $\tilde{\epsilon}^g_{j,l}$ 
(we have added the subscript $g$ to emphasize that we are dealing with 
a Gaussian field.)  

The $\tilde{\epsilon}^g_{j,l}$ are normally distributed with covariance 
given by
\begin{equation}
\langle\tilde{\epsilon}^g_{j,l}\tilde{\epsilon}^g_{j',l'}\rangle
= P_{j,l}\delta_{j,j'}\delta_{l,l'}.
\end{equation}
For a homogeneous Gaussian field, $P_{j,l}$ is $l$-independent, i.e., 
$P_{j,l}=P_j$. Thus, $P_j$ is the power spectrum of the initial field.
(Pando \& Fang 1998b, Feng \& Fang 2000, Fang \& Feng, 2000.)
Because the 2$^j$ WFCs $\tilde{\epsilon}^g_{j,l}$ form a fair ensemble,
the DWT power spectrum $P_j$ of the initial field can be calculated by
\begin{equation}
P_j=\frac{1}{2^j}\sum_{l=0}^{2^j-1}
(\tilde{\epsilon}^g_{j,l} - \langle \tilde{\epsilon}^g_{j,l} \rangle)^2,
\end{equation}
where
\begin{equation}
\langle \tilde{\epsilon}^g_{j,l} \rangle =
\frac{1}{2^j}\sum_{l=0}^{2^j-1}\tilde{\epsilon}^g_{j,l}.
\end{equation}

In our numerical calculation, we generally use the compactly supported 
basis Daubechies 4 (D4) (Daubechies 1992). The choice of 
particular compactly supported wavelet basis does not effect the 
statistical results (Pando \& Fang 1996) and the D4 wavelet is 
chosen as a matter of convenience. Although the so called Haar wavelet 
appears to be simpler, it is not well behaved in 
scale space (Fang \& Thews 1998.)    

\subsection{An example}

To demonstrate the scale-by-scale Gaussian mapping, we recover 
the initial Gaussian mass field of a simulation sample of Ly$\alpha$ forests.

The simulation samples are produced by a 
semi-analytic model of the intergalactic medium (Bi 1993; Fang  
et al. 1993.) This model approximately fits most  observed features of 
the Ly$\alpha$ clouds, including the column density 
distribution and the number density of the lines, the distribution of 
equivalent widths and their redshift dependence, the clustering of the 
lines, and even the 3rd- and 4th-order non-Gaussian features (Bi, Ge \& 
Fang 1995; Bi \& Davidsen 1997, Feng \& Fang, 2000.) We produce the 
simulation samples in redshift range 
$z=2.066\sim 2.436$ with $2^{14}$ pixels. The corresponding simulation 
size in the CDM model is 189.84 h$^{-1}$Mpc in comoving space. These 
scales weakly depend on $\Omega$.

Figure 1 shows a typical realization of the simulated Ly$\alpha$,
including the flux $F$, density contrast of IGM, $\delta_{IGM}$,
peculiar velocity of the IGM, $V_{pec}$ and the column density of
neutral hydrogen $N_{HI}$. The recovered IGM density distributions 
on scales $10/2^j$ h$^{-1}$ Mpc and $j=6, 9, 12$ are plotted in
Fig. 2. These results show that the density distributions can be 
well reconstructed by the DWT algorithm. The small deviation of the 
recovered distributions from the original one is due mainly to 
the effect of the peculiar velocity. Figure 3 plots the DWT power 
spectra of the initial and recovered fields. It shows that the 
recovered power spectrum is in excellent agreement with the initial 
one until scale $j=9$ or $k = 10$ h Mpc$^{-1}$. This means that the
effect of the peculiar velocity is significant only on small
scales $j>9$.  

However, our goal is to put the scale-of-scale algorithm on a firm foundation
and so recovering the power spectrum from simulation samples is insufficient
for our purposes. We must investigate the conditions under which the basic 
assumption of the scale-of-scale algorithm -- the order-preserving 
transformation -- is reasonable.   

\subsection{Conditions for quasi-locality}

The scale-by-scale Gaussian mapping assumes the order preservation
between the WFCs $\tilde{\epsilon}_{j,l}$ of the evolved field and the 
initial WFCs $\tilde{\epsilon}^0_{j,l}$ at each scale $j$. For scale $j$, 
index $l$ corresponds to a spatial range $L/2^j$. The scale-by-scale 
Gaussianization algorithm does not require a point-point (or pixel-pixel) 
locality, but only a locality in the range of $\Delta x = L/2^j$. 

For scale $j$, the corresponding wavenumber is $k=2\pi 2^j/L$ and we have 
\begin{equation}
\Delta x \propto 2\pi/k.
\end{equation}
That is, the size of the locality $\Delta x$ in the scale-by-scale 
Gaussian mapping is scale-adaptive. Only for wavelengths
corresponding to one pixel is pixel-pixel locality required.

Obviously, if the locality condition is valid, the mode-mode correlation
will be dominated by the local term. Thus, one can set a condition for the 
locality as
\begin{equation}
\frac {\langle \tilde\epsilon_{j,l}\tilde\epsilon_{j,l'} \rangle} 
   {\langle \tilde\epsilon^2_{j,l} \rangle} \ll 1, 
 \ \ l \neq l'.
\end{equation}
For cosmological problems, the second order correlations 
$\langle \tilde\epsilon_{j,l}\tilde\epsilon_{j,l'} \rangle$ only
depend on the difference $\Delta l = l'-l$, and eq.(9) can be 
calculated as
\begin{equation}
\frac{\kappa_{j;j}(\Delta l)}{\kappa_{j;j}(0)} \ll 1, \ \ 
   \Delta l \neq 0,
\end{equation}
where $\kappa_{j;j,}(\Delta l)$ is defined by 
\begin{equation}
\kappa_{j;j}(\Delta l) = \frac{1}{2^j} \sum_{l=0}^{2^J-1}
  \tilde\epsilon_{j,l}\tilde\epsilon_{j,l+ \Delta l}.
\end{equation} 
The ratios (10) or (11) can be used to estimate the error caused by
the locality assumption.

\section{The evolution of the DWT modes in the Zel'dovich
  approximation}

In this section, we will show that conditions eqs.(10) or (11) are
reasonable in the weakly non-linear regime. We use the Zel'dovich 
solution, which describes the growing mode of the 
gravitational clustering in the quasi-linear and even non-linear 
regime until the variance of the density field is of order unity.

\subsection{The Zel'dovich solution in the DWT representation}

In the Zel'dovich approximation [eq.(1)], the displacement vector 
field ${\bf S}({\bf q}, t)$ is given by
\begin{equation}
{\bf S}({\bf q}, t) = - b(t)\nabla \phi|_{\bf q},
\end{equation}
where $b(t)$ is the linear growth factor and $\phi({\bf q})$ is the 
initial irrotational peculiar velocity potential (Catelan 1995.) The
Eulerian density field can be described as 
\begin{equation}
\rho({\bf x},t) = \bar{\rho}(t)
\int d^3q \delta^D[{\bf x - q - S(q}, t)],
\end{equation}
where $\delta^D({\bf x})$ denotes the 3-dimensional Dirac delta function.

In the DWT representation, a density field $\rho({\bf x})$,
${\bf x}=(x_1,x_2,x_3)$, in range $0<x_i<L_i, \{i=1,2,3\}$ is described 
by the WFCs 
\begin{equation}
\tilde{\epsilon}_{\bf j,l}(t)= 
\int \rho({\bf x},t) \psi_{\bf j,l}({\bf x})d{\bf x},
\end{equation}
where the $\psi_{\bf j,l}({\bf x})$ are again an orthogonal and 
complete basis in 3-dimensions, constructed by the direct product of the
1-dimension wavelets (Daubechies 1992; Fang \& Thews 1999)
\begin{equation}
\psi_{\bf j,l}({\bf x})=\psi_{j_1,l_1}(x_1)
\psi_{j_2,l_2}(x_2)\psi_{j_3,l_3}(x_3).
\end{equation}

Substituting eqs. (13) and (15) into eq.(14), we have
\begin{equation}
\tilde{\epsilon}_{\bf j,l}(t)=\int d^3q 
\psi_{\bf j,l}[{\bf q} - b(t)\nabla \phi]. 
\end{equation}
This is the solution of the DWT mode $({\bf j,l})$ evolution in the
Zel'dovich approximation.

\subsection{Weakly non-linear evolution of the DWT modes}

In the weakly non-linear regime, one can expand eq.(16) in a power
series with respect to $b(t)$
\begin{equation}
\tilde{\epsilon}_{\bf j,l}(t)=\sum_{n=0}^{\infty}
\frac{[-b(t)]^n}{n!}\int d^3q 
(\nabla \phi \cdot \nabla)^n\psi_{\bf j,l}({\bf q}).
\end{equation}
The evolution of the DWT mode $({\bf j,l})$ can be calculated from the 
first few terms of eq.(17). Specifically, the first 4 terms give

1. The $n=0$ term is
\begin{equation}
\tilde{\epsilon}^{(0)}_{\bf j,l}(t) = \int \psi_{\bf j,l}({\bf q}) d^3q = 0,
\end{equation}
where we used the admissibility property of wavelets. This is the uniform 
background.

2. The $n=1$ term is
\begin{equation}
\tilde{\epsilon}^{(1)}_{\bf j,l}(t) = -b(t)\int d^3q
(\nabla \phi \cdot \nabla)\psi_{\bf j,l}({\bf q}) =
b(t)\int d^3q \nabla^2 \phi \psi_{\bf j,l}({\bf q}) 
\end{equation}
Because $\nabla^2 \phi$ is equal to the initial density contrast 
$\delta_0({\bf x})$, we have 
\begin{equation}
\tilde{\epsilon}^{(1)}_{\bf j,l}(t) =
b(t) \int d^3q \delta_0({\bf q}) \psi_{\bf j,l}({\bf q}) =
b(t)\tilde{\epsilon}^0_{\bf j,l}
\end{equation}
where $\tilde{\epsilon}^0_{\bf j,l}$ is given by
\begin{equation}
\tilde{\epsilon}^0_{\bf j,l}=\int d^3x\delta_0({\bf x})
\psi_{\bf j,l}({\bf x}).
\end{equation}
Therefore, $\tilde{\epsilon}^0_{\bf j,l}$ is the initial density
perturbation of mode $({\bf j,l})$. Equation (20) gives the 
linear growth of mode $({\bf j,l})$ as $b(t)$ is taken to be the 
growth solution of the density perturbations.

3. The $n=2$ term is
\begin{equation}
\tilde{\epsilon}^{(2)}_{{\bf j,l}}(t)= \frac{b^2(t)}{2}
\int d^3q (\nabla \phi \cdot \nabla)^2\psi_{{\bf j,l}}({\bf q})=
 -b^2(t)\int \nabla^2 \phi \nabla \phi \cdot \nabla\psi_{{\bf j,l}}({\bf q}).
\end{equation}
One can also decompose the initial velocity field $\nabla \phi$
into the DWT modes $({\bf j,l})$ as
\begin{equation}
\tilde{\epsilon}^{V_i}_{{\bf j,l}}=
\int d^3x \frac{\partial \phi}{\partial x_i}
\psi_{j,l}({\bf x}).
\end{equation} 
Thus, eq.(25) can be rewritten as
\begin{equation}
\tilde{\epsilon}^{(2)}_{\bf j,l}(t)= - b^2(t)\tilde{\epsilon}^0_{\bf j',l'}
\tilde{\epsilon}^{V_i}_{{\bf j'',l''}}\int d^3q\psi_{\bf j',l'}({\bf q}) 
\psi_{\bf j'',l''}({\bf q}) 
\frac{\partial}{\partial q_i}
\psi_{\bf j,l}({\bf q})
\end{equation}
where the summation conventions for ${\bf j',l'}$, ${\bf j'',l''}$ and
$i$ are employed. Since the 3-dimensional  wavelets are given by the direct 
product of  1-dimensional wavelets [eq.(15)], all terms in the r.h.s. of
eq.(27) contain the three-wavelet integrals 
$\int dx \psi_{j',l'}(x)\psi_{j'',l''}(x)\frac{d}{d x}\psi_{j,l}(x)$.
These integrals are non-zero only if the spatial ranges of the modes 
${ j', l'}$ and ${ j'',l''}$  overlap ${ j, l}$. For the Daubechies 4 wavelet, 
$ l'$ and $ l''$  should not differ from $l$ by more than 2. Thus, this 
integral requires that $(l,l',l'')$ be localized.  

4. The $n=3$ term is
\begin{eqnarray}
\tilde{\epsilon}^{(3)}_{\bf j,l}(t) & = &
-\frac{b^3(t)}{6}
\int d^3q (\nabla \phi \cdot \nabla)^3\psi_{{\bf j,l}}({\bf q}) \\ \nonumber
& = & \frac{b^3(t)}{2}\tilde{\epsilon}^0_{\bf j',l'}
\tilde{\epsilon}^{V_i}_{{\bf j'',l''}}
\tilde{\epsilon}^{V_k}_{{\bf j''',l'''}}
\int d^3q\psi_{\bf j',l'}({\bf q})\psi_{\bf j'',l''}({\bf q}) 
\psi_{\bf j''',l'''}({\bf q})\frac{\partial}{\partial q_i}
\frac{\partial}{\partial q_k}\psi_{\bf j,l}({\bf q})
\end{eqnarray}
Similar to the 3-wavelet integral, the 4-wavelet integrals in eq.(25) 
require that the modes 
${\bf j', l'}$, ${\bf j'',l''}$, ${\bf j''',l'''}$ and
${\bf j, l}$ be localized. 

Thus, the behavior of mode $({\bf j,l})$ in the weak 
non-linear regime can be estimated as
\begin{equation}
\tilde{\epsilon}_{\bf j,l}(t)=\tilde{\epsilon}^{(1)}_{\bf j,l}(t)
+\tilde{\epsilon}^{(2)}_{\bf j,l}(t)+ \tilde{\epsilon}^{(3)}_{\bf j,l}(t).
\end{equation}
Considering the locality of the wavelet integrals in eqs(24) and (25),
eq.(26) yields
\begin{eqnarray}
\tilde{\epsilon}_{\bf j,l}(t) & \simeq &
b(t)\tilde{\epsilon}^0_{\bf j,l} \left [1 
  - b(t)\tilde{\epsilon}^{V_i}_{{\bf j'',l''}}
\int d^3q\psi_{\bf j,l}({\bf q})\psi_{\bf j'',l''}({\bf q})
\frac{\partial}{\partial q_i}
\psi_{\bf j,l}({\bf q}) \right . \\ \nonumber  
 & & \left . + \frac{b^2(t)}{2}
\tilde{\epsilon}^{V_i}_{{\bf j'',l''}}
\tilde{\epsilon}^{V_k}_{{\bf j''',l'''}}
\int d^3q\psi_{\bf j,l}({\bf q})\psi_{\bf j'',l''}({\bf q}) 
\psi_{\bf j''',l'''}({\bf q})\frac{\partial}{\partial q_i}
\frac{\partial}{\partial q_k}\psi_{\bf j,l}({\bf q}) \right. ]
\end{eqnarray}

If the initial density and velocity perturbations are Gaussian, we have
\begin{eqnarray}
\langle \tilde{\epsilon}^0_{\bf j,l}\rangle=
  \langle \tilde{\epsilon}^{V_i}_{\bf j,l}\rangle =0 \\ \nonumber
\langle \tilde{\epsilon}^{V_i}_{\bf j',l'}
\tilde{\epsilon}^0_{\bf j'',l''}\rangle = 0 
\end{eqnarray} 
and 
\begin{eqnarray}
\langle \tilde{\epsilon}^0_{\bf j',l'}\tilde{\epsilon}^0_{\bf j'',l''} 
\rangle 
 = \langle \tilde{\epsilon}^{V_i}_{\bf j'',l''}
\tilde{\epsilon}^{V_k}_{\bf j''',l'''}\rangle = 0, \\ \nonumber
\ \ \ \ \  
{\rm if}\ {\bf j'} \neq {\bf j''}\ {\rm or}\ {\bf l'} \neq {\bf l''}.
\end{eqnarray}
All high order cumulants of $\tilde{\epsilon}^0_{\bf j",l"}$
and $\tilde{\epsilon}^{V_i}_{\bf j',l'}$ are also zero.

Thus, in the bracketed expression of the r.h.s. of eq.(27), the second 
term is zero on average because the initial Gaussian velocity perturbations 
$\langle\tilde{\epsilon}^{V_i}_{{\bf j'',l''}} \rangle =0$.
The third term in the bracket is proportional to the variance of 
$\tilde{\epsilon}^{V_i}_{{\bf j,l}}$, and then to the variance of 
$\tilde{\epsilon}^{0}_{{\bf j,l}}$.
Therefore, large initial WFCs $\tilde{\epsilon}^0_{\bf j,l}$ statistically
evolve into large $\tilde{\epsilon}_{\bf j,l}(t)$, and low initial  
WFCs into low WFCs. The evolution is monotonic. 

\subsection{Quasi-locality of second order correlations in phase space}

The locality can be expressed by the mode-mode correlations. For the second
order correlation between modes $({\bf j^1,l^1})$ and 
$({\bf j^2,l^2})$, the non-zero terms up to order $b^4$ are
\begin{eqnarray}
\lefteqn{ \langle \tilde{\epsilon}_{\bf j^1,l^1}(t)
\tilde{\epsilon}_{\bf j^2,l^2}(t)\rangle  =  }  \\ \nonumber
 & = &
\langle \tilde{\epsilon}^{(1)}_{\bf j^1,l^1}(t)
\tilde{\epsilon}^{(1)}_{\bf j^2,l^2}(t)\rangle
+ \langle \tilde{\epsilon}^{(2)}_{\bf j^1,l^1}(t)
\tilde{\epsilon}^{(2)}_{\bf j^2,l^2}(t)\rangle
+ \langle \tilde{\epsilon}^{(1)}_{\bf j^1,l^1}(t)
\tilde{\epsilon}^{(3)}_{\bf j^2,l^2}(t)\rangle
+
\langle \tilde{\epsilon}^{(3)}_{\bf j^1,l^1}(t)
\tilde{\epsilon}^{(1)}_{\bf j^2,l^2}(t)\rangle  \\ \nonumber
& = & b^2(t)\langle \tilde{\epsilon}^0_{\bf j^1,l^1}
\tilde{\epsilon}^0_{\bf j^2,l^2} \rangle 
 +b^4(t)\langle \tilde{\epsilon}^0_{\bf j,l}
\tilde{\epsilon}^0_{\bf j',l'}
\tilde{\epsilon}^{V_i}_{{\bf j'',l''}}
\tilde{\epsilon}^{V_k}_{{\bf j''',l'''}}  \rangle \times \\ \nonumber
& & \ \ \ \int d^3q\psi_{\bf j,l}({\bf q}) 
\psi_{\bf j'',l''}({\bf q}) 
\frac{\partial}{\partial q_i}
\psi_{\bf j^1,l^1}({\bf q})
\int d^3q\psi_{\bf j',l'}({\bf q}) 
\psi_{\bf j''',l'''}({\bf q}) 
\frac{\partial}{\partial q_i}
\psi_{\bf j^2,l^2}({\bf q})      \\ \nonumber 
 &  &  +\frac{b^4(t)}{2} \left\{ \langle \tilde{\epsilon}^0_{\bf j^1,l^1}
\tilde{\epsilon}^0_{\bf j',l'}
\tilde{\epsilon}^{V_i}_{{\bf j'',l''}}
\tilde{\epsilon}^{V_k}_{{\bf j''',l'''}}\rangle 
 \int d^3q\psi_{\bf j',l'}({\bf q})\psi_{\bf j'',l''}({\bf q}) 
\psi_{\bf j''',l'''}({\bf q})\frac{\partial}{\partial q_i}
\frac{ \partial}{\partial q_k}\psi_{\bf j^2,l^2}({\bf q}) \right. 
   \\ \nonumber
& & \left. + [({\bf j^1,l^1}) \rightleftharpoons ({\bf j^2,l^2})] 
  \right\}.
\end{eqnarray}
The term $b^2(t)$ is the linear term and the $b^4(t)$ term gives the first 
non-linear correction.

Due to the locality of the 3- and 4-wavelet integrals, eq.(30) yields
\begin{equation}
\langle \tilde{\epsilon}_{\bf j^1,l^1}(t)
\tilde{\epsilon}_{\bf j^2,l^2}(t)\rangle \simeq 0.
\end{equation}
The second order correlations between the DWT modes 
$({\bf j,l})$ are localized [eqs.(9) and (10)].

Since the 3- and 4-wavelet integrals do not require that scales ${\bf j}$ 
be the same as ${\bf j'}$, ${\bf j''}$, and ${\bf j'''}$, there exist 
correlations between modes on different scales but localized in the 
same spatial range. That is
\begin{equation}
\langle \tilde{\epsilon}_{\bf j^1,l^1}
\tilde{\epsilon}_{\bf j^2,l^1}\rangle \neq 0.
\end{equation}
This second order correlation actually is a non-Gaussian correlation 
which corresponds to the non-random phase of the Fourier modes,
or the phase-phase correlations between the Fourier modes 
(Feng \& Fang 2000)

Moreover, the spatial size of a mode on scale $j$ is $L/2^j$. 
Thus, the scale-scale correlations between $j'$ and $j$ lead to 
a dependence of perturbations at $L/2^{j'}$ on the perturbations 
at $L/2^j$. The two perturbations are not perfectly localized in
the same area, and therefore, the coupling between the DWT modes 
is quasi-local. In the fully developed non-linear regime, the 
strong scale-scale correlations will finally lead to non-local 
correlations of the DWT modes. 

\subsection{Non-Gaussianity of third-order correlations}

The non-Gaussianities on higher-order correlations caused by the 
weakly non-linear evolution can be calculated by the cumulants of 
$\tilde{\epsilon}_{\bf j,l}$. Using eq.(29), the third order 
correlation up to order $b^4$ is
\begin{eqnarray}
\langle \tilde{\epsilon}_{\bf j^1,l^1}
\tilde{\epsilon}_{\bf j^2,l^2}
\tilde{\epsilon}_{\bf j^3,l^3} \rangle & = & 
\langle \tilde{\epsilon}^{(1)}_{\bf j^1,l^1}(t)
\tilde{\epsilon}^{(1)}_{\bf j^2,l^2}(t)
\tilde{\epsilon}^{(1)}_{\bf j^3,l^3}(t)\rangle   \\ \nonumber 
& & + [\langle \tilde{\epsilon}^{(1)}_{\bf j^1,l^1}(t)
\tilde{\epsilon}^{(1)}_{\bf j^2,l^2}(t)
\tilde{\epsilon}^{(2)}_{\bf j^3,l^3}(t)\rangle
+ {\rm \ 2 \ terms \ with \ cyc. \ permutations}].
\end{eqnarray}
For Gaussian initial perturbation eq.(28), we have
\begin{equation} 
\langle \tilde{\epsilon}^{(1)}_{\bf j^1,l^1}(t)
\tilde{\epsilon}^{(1)}_{\bf j^2,l^2}(t)
\tilde{\epsilon}^{(1)}_{\bf j^3,l^3}(t) \rangle \propto 
\langle \tilde{\epsilon}^{0}_{\bf j^1,l^1}(t)
\tilde{\epsilon}^{0}_{\bf j^2,l^2}(t)
\tilde{\epsilon}^{0}_{\bf j^3,l^3}(t)\rangle =0.
\end{equation}
Using eq.(24) we have
\begin{eqnarray}
\lefteqn {\langle \tilde{\epsilon}^{(1)}_{\bf j^1,l^1}(t)
\tilde{\epsilon}^{(1)}_{\bf j^2,l^2}(t)
\tilde{\epsilon}^{(2)}_{\bf j^3,l^3}(t)\rangle
   \propto } \\ \nonumber 
  & & \langle \tilde{\epsilon}^{0}_{\bf j^1,l^1}(t)
\tilde{\epsilon}^{0}_{\bf j^2,l^2}(t)
\rangle\tilde{\epsilon}^0_{\bf j',l'}
\tilde{\epsilon}^{V_i}_{{\bf j'',l''}}\rangle
\int d^3q\psi_{\bf j',l'}({\bf q}) 
\psi_{\bf j'',l''}({\bf q})\frac{\partial}{\partial q_i}
\psi_{\bf j^3,l^3}({\bf q}) =0.
\end{eqnarray}
Therefore, the largest two terms of eq.(33) are zero and we have 
\begin{equation}
\langle \tilde{\epsilon}_{\bf j^1,l^1}
\tilde{\epsilon}_{\bf j^2,l^2}
\tilde{\epsilon}_{\bf j^3,l^3} \rangle \simeq 0.
\end{equation}
That is, the one-point distribution of the DWT 
modes $({\bf j,l})$ contains at most a small skewness.

This result can also be obtained from the hierarchical clustering
or linked-pair approximation as
\begin{eqnarray}
\lefteqn {\langle \delta({\bf x^1})\delta({\bf x^2})\delta({\bf x^3}) \rangle 
 \simeq Q_3
[\langle \delta({\bf x^1})\delta({\bf x^2})\rangle \langle 
\delta({\bf x^1})\delta({\bf x^3}) \rangle } \\ \nonumber
 & & + {\rm \ 2 \ terms \ with \ cyc. \ permutations}],
\end{eqnarray}
where $\delta({\bf x})=[\rho({\bf x})-\bar{\rho}]/\bar{\rho}$. 
Since wavelet $\psi_{j,l}$ is admissible, i.e., 
$\int \psi_{j,l}(x)dx=0$, eq.(17) gives
\begin{equation}
\tilde{\epsilon}_{\bf j,l}(t)= \frac{1}{\bar{\rho}}
\int \delta({\bf x}) \psi_{\bf j,l}({\bf x})d{\bf x}.
\end{equation}
We can take $\bar{\rho}=1$, i.e., the mean density is normalized.

Expressing eq.(38) in the wavelet basis 
$\psi_{\bf j,l^1}({\bf x^1})\psi_{\bf j^2,l^2}({\bf x^2})
\psi_{\bf j^3,l^3}({\bf x^3})$, we have 
\begin{eqnarray}
\lefteqn{\langle \tilde{\epsilon}_{\bf j,l^1}
\tilde{\epsilon}_{\bf j,l^2}
\tilde{\epsilon}_{\bf j,l^3} \rangle \simeq 
Q_3 \sum_{\bf l',l''}a^3_{\bf l^1, l^{'1},l^{''1}}
  [\langle \tilde{\epsilon}_{\bf j,l^{'1}}
\tilde{\epsilon}_{\bf j,l^2}\rangle
\langle\tilde{\epsilon}_{\bf j,l^{''1}}
\tilde{\epsilon}_{\bf j,l^3} \rangle } \\ \nonumber 
 & & + {\rm \ 3 \ terms \ with \ cyc. \ permutations}].
\end{eqnarray}
where $a^3_{\bf l^1, l^{'1},l^{''1}}$ is given by the 3-wavelet integral, 
\begin{equation}
a^3_{\bf l^1, l^{'1},l^{''1}}
=\int \psi_{\bf j,l^1}({\bf x})\psi_{\bf j,l^{'1}}({\bf x})
  \psi_{\bf j,l^{''1}}({\bf x}) d{\bf x}.
\end{equation}
$a^3_{\bf l^1, l^{'1},l{''1}}$ generally is small and exactly equal to 
zero for the Haar wavelets (Mallat 1989a; Meyer 1992). Therefore, a 
hierarchical clustered field is only weakly skewed. Note however, that 
this does {\it not} imply that the skewness in the DWT representation 
is small in general.  

\subsection{Non-Gaussianity of fourth-order correlations}

In contrast, the 4th-order correlations of the WFCs given by eq.(26) 
are generally non-zero at order $b^4$. The first significant non-zero 
higher order term not caused by $\rho \ge 0$ is the kurtosis defined by
\begin{equation}
K_j=\frac {\langle (\tilde{\epsilon}_{\bf j,l}-
        \langle\tilde{\epsilon}_{\bf j,l}\rangle)^4\rangle}
     {\langle\tilde{\epsilon}^2_{\bf j,l}\rangle^2}-3.
\end{equation}
The kurtosis can be considered as a special case ($j=j'$ and $\Delta l=0$)
of the fourth order scale-scale correlations defined by
\begin{equation}
C^{2,2}_{j,j'}(\Delta l) = \frac{2^{j'}\sum_{l=0}^{2^{j'}-1}
\tilde{\epsilon}^2_{j;[l/2^{j'-j}]+\Delta l} \;
\tilde{\epsilon}^2_{j';l}} {\sum \tilde{\epsilon}^2_{j,[l/2^{j'-j}]}
\sum \tilde{\epsilon}^2_{j';l}}.
\end{equation}
$C^{2,2}_{j;j'}(\Delta l)$ measures the
correlations between the perturbations of modes on scales $j$ and $j'$, at
two positions separated a distance $\Delta l L/2^j$. In the case of 
$\Delta l=0$, $C^{2,2}_{j;j'}(0)$ measures the correlation between 
fluctuations on scale $j$ and $j'$ at the {\it same} physical point.

Using the locality of the 4-wavelet integral in eq.(25),
one can show that the scale-scale correlations in the weakly non-linear 
regime are also localized, i.e.,
\begin{equation}
C^{2,2}_{j;j'}(\Delta l \neq 0) \ll C^{2,2}_{j;j'}(0).
\end{equation}

In the weakly non-linear regime, the phase space behavior of 
gravitational clustering for an initially Gaussian mass field can be 
summarized as
\begin{enumerate}
\item The evolution is spatially quasi-localized, i.e., the correlations
  between WFCs at different positions are always less than that at the 
  same position.
\item The density field contains a very small skewness.
\item The major non-Gaussianities are local scale-scale correlations.
\end{enumerate}

It should be pointed out that if a distribution possesses the
above-listed features, it does not mean that the initial mass field is
Gaussian. Actually, for a distribution to exhibit these features, we only 
need the following conditions on the initial density perturbation:
\begin{equation}
\langle \tilde{\epsilon}^0_{\bf j,l}\tilde{\epsilon}^0_{\bf j',l'}
\rangle = 0,
\end{equation}
when the spatial range of ${\bf l}$ doesn't overlap  ${\bf l'}$,
and
\begin{equation}
\langle \tilde{\epsilon}^0_{\bf j,l}\tilde{\epsilon}^0_{\bf j',l'}
\rangle \neq 0,
\end{equation}
when the spatial range of ${\bf l}$ does overlap ${\bf l'}$.
The conditions given by eqs.(44) and (45) are weaker than eqs.(28)
and (29). The initial field given by eqs.(44) and (45) can be 
non-Gaussian.

Equations (44) and (45) show once again that the statistical properties
of a random field should be described by two types of correlations: one
with respect to scale $j$ and the other with respect to position $l$.
These two types of correlations correspond to correlations with 
respect to the phase and amplitude in the Fourier representation. 
It is, in fact, possible to construct clustering models which lead to a 
density field with a Poisson or Gaussian PDF with respect to $l$, but 
that are highly scale-scale correlated (Greiner, Lipa \& Carruthers, 1995.)

\subsection{Redshift space}

If the density field is viewed in redshift space, the observed radial 
position is given by the radial velocity consisting of the uniform Hubble 
flow and the peculiar motion ${\bf v}({\bf x})$. Thus the position 
${\bf x}$ in eq.(1) should be replaced by
\begin{equation}
{\bf s}={\bf x} + [\hat{\bf z}\cdot{\bf v}({\bf x})]\hat{\bf z},
\end{equation}
where $\hat{\bf z}$ is taken to be in the direction of the line of sight.
The second term in eq.(46) is the correction due to the radial peculiar
velocity. 

In redshift space the Zel'dovich solution eq.(16) is
replaced by (e.g. Taylor \& Hamilton 1996)
\begin{equation}
\rho({\bf s},t) = \bar{\rho}(t)
\int d^3q \delta^D[{\bf s - q - S^s(q}, t)].
\end{equation}
In the linear approximation of the velocity field, the displacement vector
field is 
\begin{equation}
{\bf S^s}( {\bf q}) ={\bf S}({\bf q})+
  \Omega^{0.6}[\hat{\bf z}\cdot {\bf S}({\bf x})]\hat{\bf z},
\end{equation}
where $\Omega$ is the cosmological density parameter. Thus, the solution
[eq.(19)] becomes
\begin{equation}
\tilde{\epsilon}_{\bf j,l}(t)=\int d^3q 
\psi_{\bf j,l}[{\bf q} +  {\bf S^s(q}, t)]. 
\end{equation}
This is the Zel'dovich solution for the DWT modes in redshift space. 
Mathematically, eq.(48) is equivalent to eq.(16), with
${\bf S^s(q}, t)$ replacing ${\bf S(q}, t)$.  Therefore, all results 
drawn from eq.(16) still hold for eq.(49) if the intersection of the 
particle trajectories has not yet happen. 

It is known that the non-linear effect of the redshift distortion,
like the Fingers-of-God in the galaxy distribution in redshift space, cannot 
be modeled by the Zel'dovich solution. In other words, the non-linear 
effects of the redshift distortion are not negligible if the Zel'dovich
solution fails. Thus, the scale on which the quasi-locality assumption is
no longer correct is also the scale where the non-linear redshift distortion 
effects emerge. This result is consistent with
the numerical example shown in \S 2.3, which shows that on scales as small 
as $j=9$, the scale-by-scale reconstruction is still very good, and the 
redshift distortion effects are not yet significant. 

\section{Quasi-locality in the Ly$\alpha$ forests}

To test the predictions on the quasi-locality of the DWT mode-mode 
correlations (\S 3.5), we analyze samples of the QSO Ly$\alpha$ absorption 
spectrum which are believed to trace the underlying mass field in
the weakly non-linear regime. The second point listed in \S 3.5 has been 
tested by our previous studies, which showed no skewness above a 
positive definite random distribution for the distributions of Ly$\alpha$ 
forests lines (Pando \& Fang 1998a) or the transmitted flux 
(Feng \& Fang 2000.) In this section, we will focus on the points 1 and 3.

The QSO Ly$\alpha$ absorption spectrum samples used in our analysis 
are: 1.) the transmitted flux of HS1700+64 Ly$\alpha$ absorptions; 2.) 
the Ly$\alpha$ forest lines from moderate resolution spectrum, including 
a sample compiled by Lu, Wolfe, \& Turnshek (LWT) (1991) that contains 
$\sim$ 950 lines from the spectra of 38 QSOs, and a sample compiled by 
Bechtold (JB) (1994) which contains a total $\sim$ 2800 lines from 78 QSO's 
spectra in which 34 high-redshift QSO's were observed, and 3.) the 
Ly$\alpha$ forest lines from high resolution spectrum, including 
the data of Hu et al. (HU) (1995) which has $\sim$ 1056 lines from four 
quasars in wavelength range 4300-5100 \AA$\,$ measured with the HIRES 
spectrograph on the Keck, and the HIRES data of QSO HS1946+7658 
0f Kirkman \& Tytler (HT) (1997).

The simulation samples used in this section are the same ones used in
\S 2.3.

\subsection{Quasi-locality of the second order DWT mode-mode correlation}

Figure 4 shows $\kappa_{j,j}(\Delta l)$ for 100 realizations of the 
simulated Ly$\alpha$ transmitted flux and QSO HS1700+64. Here we 
use the Daubechies 4 wavelets in our calculations as these 
wavelets are better behaved in scale space.
Figure 4 shows clearly that the condition for locality eq.(13)  
holds. The correlations  $\kappa_{j,j}(\Delta l)$ for all 
$\Delta l\neq 0$ are sharply lower than $\kappa_{j,j}(0)$. For small 
scales $j> 5$, at the 95\% confidence level, the non-local correlations are 
no more than 20\%.  

Figure 5 shows the correlation $\kappa_{j,j+1}(\Delta l)$, 
the second order correlations between scales $j$ and $j+1$. Comparing
with $\kappa_{j,j}(0)$, the second order non-local scale-scale 
correlations ($\Delta l \neq 0$) show no power. For the local case,
$\kappa_{j,j+1}(0)$ generally also has no power. But it is interesting to
note that $\kappa_{j,j+1}(0)$ becomes significant on very small scales 
$j = 9, \ 10.$  This is a second order non-Gaussian detection. 
One can similarly detect other second order scale-scale correlations, 
such $(j,j+2)$, $(j, j+3)$ etc. with similar results.

Figures 6 and 7 are similar to Fig. 4 but for the LWT/JB and HU/KT Ly$\alpha$ 
forest line samples. Although the identification of 
absorption lines from transmitted flux may introduce an arbitrary bias,
the LWT/JB and HU/KT data show the same behavior as
$\kappa_{j,j}(\Delta l)$ in Fig. 4. For scales  $j>5$, the locality
eq. (10) is a very good approximation, even considering the large
error in the data. For each QSO, we calculate the statistical 
quantity $\kappa_{j,j+1}(\Delta l)$, and the mean and 1-$\sigma$ error
bars are found from the ensemble of these QSOs. Since there are more QSO's 
in the LWT/JB data, the error bars for the LWT/JB data are smaller than
the high resolution data HU/KT.

Figures 8 and 9 are similar to Fig. 5 but again for the Ly$\alpha$ forest 
lines. The results once again 
show that the behavior of the second order scale-scale correlations 
for the forest line samples is the same as the
transmitted flux samples. There is no power in these second 
order scale-scale correlations.

\subsection{Quasi-locality of 4th order non-Gaussianity}

In the previous section, we detected the local scale-scale
correlation from the second order statistics. The local 
correlation of the 4th-order statistics is more prominent. In our previous
studies, we have shown that the Ly$\alpha$ forests are significantly
non-Gaussian at 4th-order, including the kurtosis (Pando \&
Fang 1998a) and the scale-scale correlations (Pando et al. 1998, Feng \& Fang
2000). In this section, we will show that although the 4th-order local
correlations are strong, the non-local correlations are weak. 

The 4th-order statistic in Figure 10 is defined by
\begin{equation}
Q^{2,2}_{j}(\Delta l) = \frac{2^{j}\sum_{l=0}^{2^{j}-1}
\tilde{\epsilon}^2_{j;l} \; \tilde{\epsilon}^2_{j;l+\Delta l}}
{\sum \tilde{\epsilon}^2_{j,l} \sum \tilde{\epsilon}^2_{j;l+\Delta
l}}.
\end{equation}
For the local case, i.e., $\Delta l=0$, $Q^{2,2}_{j}(0)$ is 
the compact form of the kurtosis
\begin{equation}
K_j = \frac{2^{j}\sum_{l=0}^{2^{j}-1}\tilde{\epsilon}^4_{j;l}}
{[\sum_{l=0}^{2^{j}-1} \tilde{\epsilon}^2_{j,l}]^2}.
\end{equation}

Figure 10 shows that the non-local correlation $Q^{2,2}_{j}(\Delta l)$ 
with $\Delta l\neq 0$ of the transmitted flux generally is around 1, 
i.e., no correlation.
On small scales $j \geq5$ the local correlation $Q^{2,2}_{j}(0)$ 
gradually becomes significant, while the non-local correlation
$Q^{2,2}_{j}(\Delta l)$ with $\Delta l\neq 0$ still remains at 1.
This result supports the quasi-locality condition.

Figure 11 displays $C^{2,2}_{j}(\Delta l)$, which is the scale-scale 
correlation between scales $j$ and $j+1$, and shows that the non-local 
scale-scale correlation $C^{2,2}_{j}(\Delta l)$ with 
$\Delta l\neq 0$ generally is around 1, i.e., no correlation. When 
the local scale-scale correlation becomes significant on scales 
$j\geq 7$, the non-local correlation
$C^{2,2}_{j}(\Delta l)$ with $\Delta l\neq 0$ still remains at 1.

Unlike the second order scale-scale correlations, the 4th-order
correlations $Q^{2,2}_{j}(0)$ and $C^{2,2}_{j}(0)$ are significant. 
There is substantial power on all scales $j \geq 5$. 

Figures 12 and 13 show the scale-scale correlations for the LWT/JB 
and HU/KT Ly$\alpha$ forest lines, respectively. Figure 9
clearly shows a steady decrease in the scale-scale amplitudes with
$\Delta l$. This again demonstrates the quasi-locality. Fig. 10 also
shows a decrease in the scale-scale amplitudes with $\Delta l$, 
but it is rather slow. Even when $\Delta l=3$, the correlation remains. 
These non-local correlations may be caused by problems in identifying the 
lines, and if so, indicates that the transmitted flux is better suited than
the Ly$\alpha$ forest lines for a Gaussianization reconstruction.

\section{Conclusions and discussions}

In this work we have demonstrated that the assumptions necessary for
reconstructing an initial density field by a scale-dependent Gaussianization
procedure are justified. The assumptions, namely the locality and 
monotonic nature of the gravitational clustering, are borne out by both 
the Zel'dovich approximation and correlation analysis of QSO Ly$\alpha$ 
samples. That is, the DWT mode-mode coupling caused by the weakly 
non-linear evolution of the gravitational clustering is quasi-local and 
that the large initial WFCs evolve into large WFCs of the observed density 
field. This result provides a solid basis for the 
DWT scale-by-scale Gaussianization reconstruction of Ly$\alpha$ forests. 

The non-Gaussianities produced by the mode-mode coupling 
during the weakly non-linear regime are mainly local scale-scale 
correlations. This result explains why the Gaussianization 
reconstruction is substantially improved by removing the local 
scale-scale correlations (Feng \& Fang 2000.) Clearing the local 
scale-scale correlation is a key for an effective Gaussianization 
reconstruction. All these features provide a solid basis for the DWT 
scale-by-scale (or scale-adaptive) Gaussianization reconstruction 
algorithm.

Since galaxies undergo a highly non-linear evolution, and their bias might 
be non-monotonic, the validity of the order preserving assumption 
for the galaxy field needs further study . Whether a sample is suitable for 
the DWT scale-by-scale reconstruction can be judged by detecting the 
non-local mode-mode correlations. If the scale-scale correlations are 
significantly non-local, the sample will not be a good candidate for a 
DWT scale-by-scale Gaussianization. The local scale-scale correlation have 
been detected in recent years for galaxy samples, such as the APM bright 
galaxy catalog (Feng, Deng \& Fang 2000.) However, it remains to
be determined whether these couplings are non-localized. 

In essence, the DWT analysis looks at the phase space behavior of 
the gravitational clustering.  While there are many 
perturbation calculations of the clustering dynamics in the Fourier 
and coordinate representations (Buchert 1993; Bouchet et al. 1995; 
Catelan 1995), the locality cannot be revealed with these calculations,
because in the Fourier representation, the phases hold the position 
information (locality) of the relevant Fourier modes. This makes it 
practically impossible to determine the locality of clustering in the 
Fourier representation. The DWT analysis may also be useful in studying 
other clustering features in the Zel'dovich approximation for which a 
phase space description is essential, such as the breakdown of the 
Galilean invariance or the lack of momentum conservation (Polyakov 1995; 
Scoccimarro 1998.)

\acknowledgments

This work was supported in part by the LWL foundation. We thank 
Dr. D. Tytler for kindly providing the data of the Keck spectrum HS1700+64. 
LLF acknowledges support from the National Science Foundation of
China (NSFC) and World Laboratory Scholarship.

\newpage

\begin{figure}
\caption{A simulated sample of the Ly$\alpha$ forest, including the flux
$F$, the IGM density contrast $\delta_{IGM}$, peculiar velocity $V_{pec}$ 
and column density of neutral hydrogen $N_{HI}$.  
}
\end{figure} 

\begin{figure}
\caption{The IGM density field (left panel) and flux (right panel). 
The solid lines are for the reconstructed density field and flux by 
the scale-by-scale Gaussianization. The dotted lines are the original 
fields.
}
\end{figure}

\begin{figure}
\caption{The DWT power spectra of the mass fields recovered by the
scale-by-scale algorithm. The error bars are the $1\sigma$ deviation
calculated for the 100 realizations. 
}
\end{figure}

\begin{figure}
\caption{The dependence of $\kappa_{j,j}(\Delta l)$ on $\Delta l$.
$\kappa_{j,j}(\Delta l)$ is normalized to $\kappa_{j,j}(0) =1$,
and the spatial distance between the two modes $(j,l)$ and 
$(j,l')$ is $D=L \Delta l$, where $L$ is the length of the sample.
The results are obtained from 100 realizations of simulated Ly$\alpha$ 
transmitted flux. The central line is the mean and the upper and 
lower lines show the 95\% confidence level. The result 
for QSO HS 1700+64 is displayed as the solid squares.}
\end{figure}

\begin{figure}
\caption{The $\Delta l$-dependence of correlation $\kappa_{j,j+1}(\Delta l)$.
where $\kappa_{j,j+1}(\Delta l)$ is normalized to $\kappa_{j,j}(0) =1$,
and the spatial distance between the two modes $(j,l)$ and 
$(j+1,l')$ is $D=L \Delta l$, where $L$ is the length of the sample.
The results are obtained from 100 realizations of simulated Ly$\alpha$ 
transmitted flux. The central line is the mean and the upper and lower
lines show the 95\% confidence level. The result for QSO 
HS1700+64 is displayed as the solid squares.
}
\end{figure}

\begin{figure}
\caption{$\kappa_{j; j}(\Delta l)$ for the moderate resolution data of
the Ly$\alpha$ forest lines LWT/JB. The case of $l = l'$, 
or $\Delta l=0$, is the localized case, and $l'+l+1$ and $l'=l+2$ show
the non-local correlations. $\kappa_{j; j}(\Delta l)$
is calculated for each QSO absorption line system and the 1-$\sigma$ bars
are determined from the variance of the data consisting of all the systems.}
\end{figure}

\begin{figure}
\caption{Same as figure 4, but for the HU/KT high resolution absorption 
samples of Ly$\alpha$ forest lines.}
\end{figure}  

\begin{figure}
\caption{The correlation $\kappa_{j,j+1}(\Delta l)$ vs. $\Delta l=l'-l$
for the moderate resolution data of Ly$\alpha$ forest lines LWT/JB. 
As expected, there is no 
power in this second order scale-scale correlation at any scale.
$\kappa_{j; j+1}(\Delta l)$
is calculated for each QSO absorption line system and the 1-$\sigma$ bars
determined from the variance of the data consisting of all the systems.
}
\end{figure}

\begin{figure}
\caption{Same as figure 5 but now using the high resolution data 
HU/KT.}
\end{figure}

\begin{figure}
\caption{The 4th-order correlation $Q_{j,j}^{2,2}(\Delta l)$ from 100 
realizations of simulated Ly$\alpha$ transmitted flux and 
QSO HS1700+64.  The spatial distance between the two modes $(j,l)$ 
and $(j,l')$ is $D=L \Delta l$, where $L$ is the length of the sample.
The central line is the mean and the upper and lower lines show 
95\% confidence level of the simulation samples. The result 
for the QSO HS1700+64 is displayed as the solid squares.
}
\end{figure}

\begin{figure}
\caption{The 4th-order scale-scale correlation 
$C_{j}^{2,2}(\Delta l)$ from 100 realizations of simulated 
Ly$\alpha$ transmitted flux and QSO HS1700+64. 
The spatial distance between the two modes $(j,l)$  and $(j+1,l')$ is 
$D=L \Delta l$, where $L$ is the length of the sample. The central
line is the mean and the upper and lower lines show the 
95\% confidence level of the simulation samples. The result for the 
QSO HS1700+64 is displayed as the solid squares.
}
\end{figure}

\begin{figure}
\caption{The 4th-order scale-scale correlations, $C_{j}^{2,2}(\Delta l)$,
for the LWT/JB data.
$C_{j}^{2,2}(\Delta l)$ is calculated for each QSO absorption
line system and the 2-$\sigma$ bars
determined from the variance of the data consisting of all the systems.
 }
\end{figure}

\begin{figure}
\caption{The 4th-order scale-scale correlations, $C_{j}^{2,2}(\Delta l)$
for the HU/KT data. $C_{j}^{2,2}(\Delta l)$ is calculated
for each QSO absorption line system and the 2-$\sigma$ bars
determined from the variance of the data consisting of all the systems.
}
\end{figure}

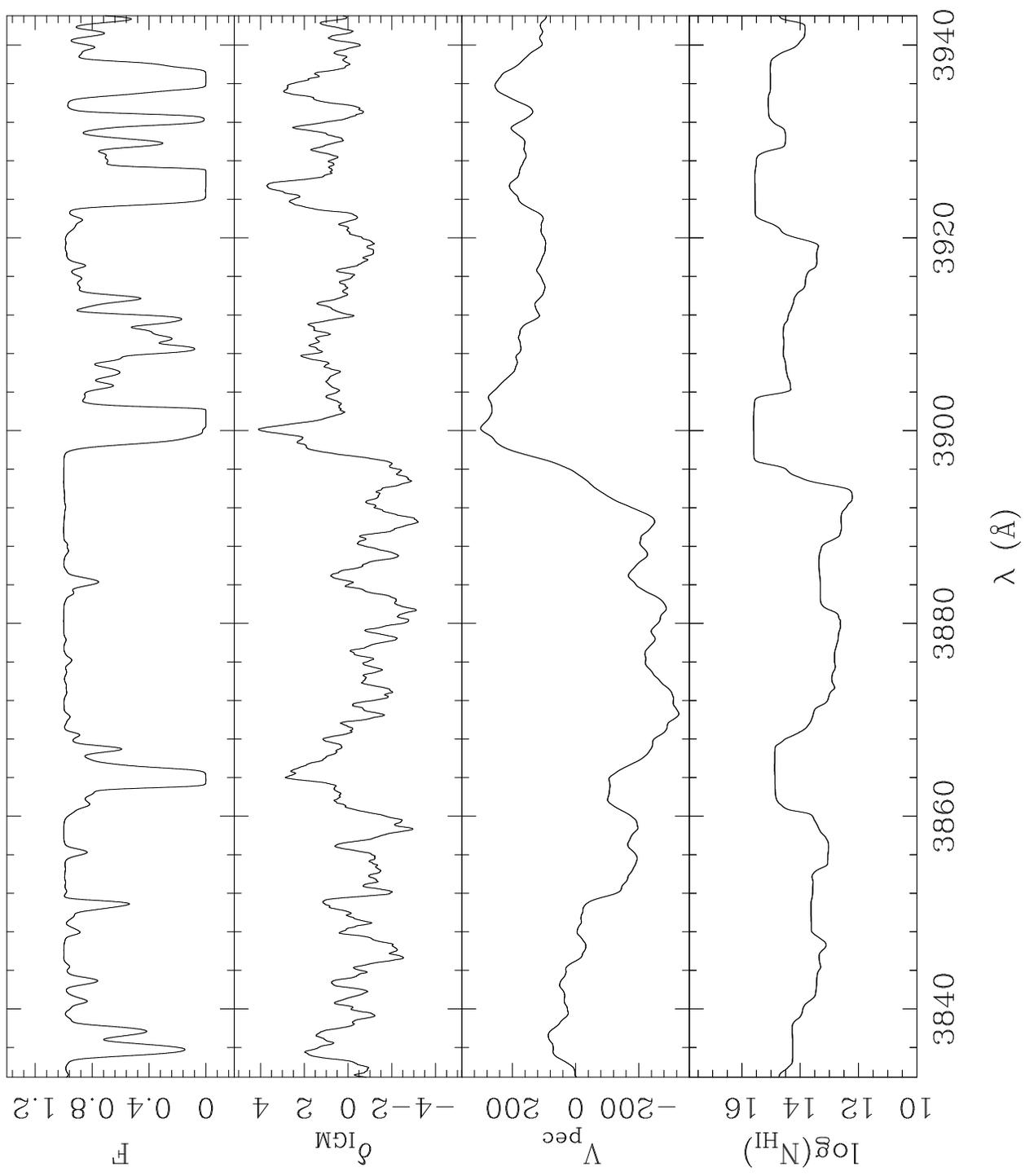
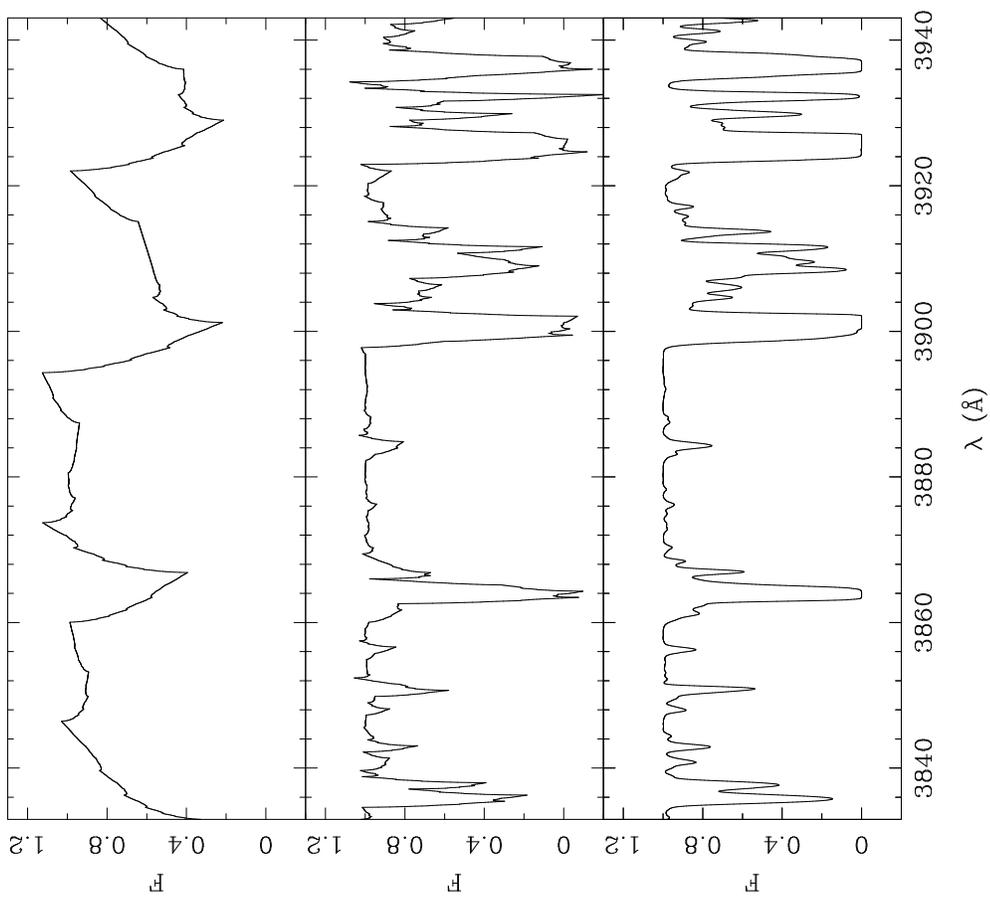
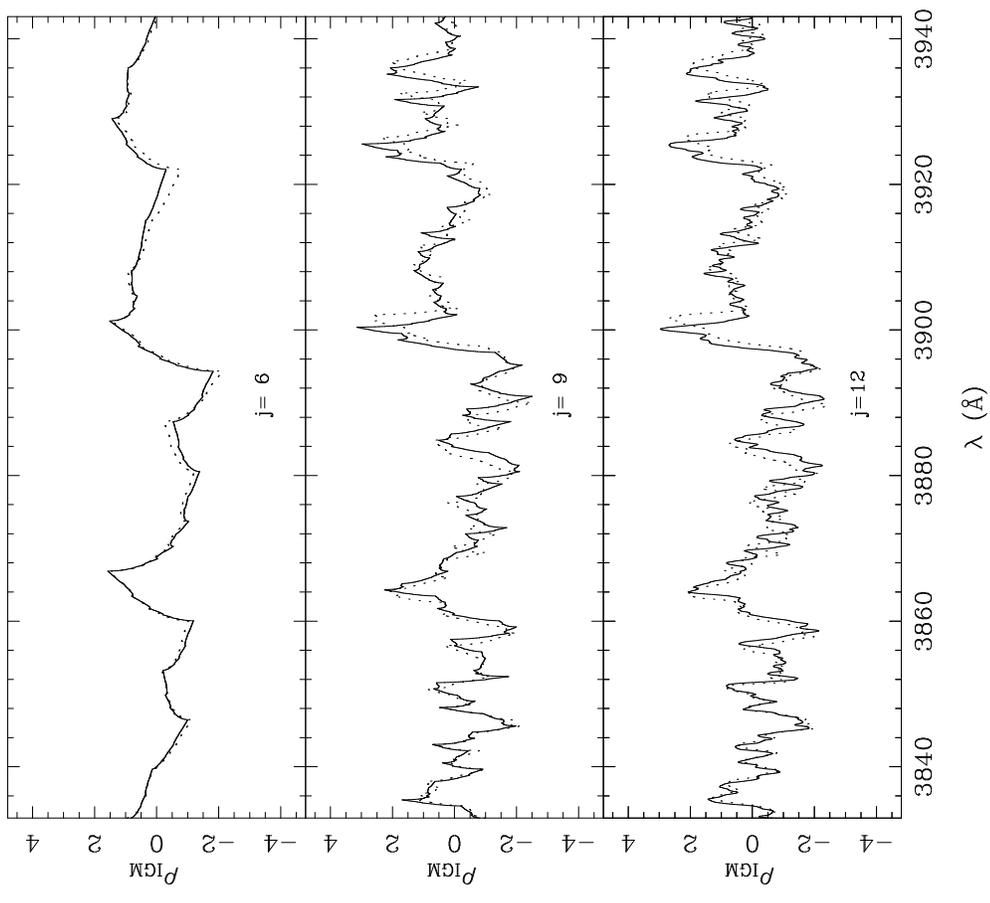
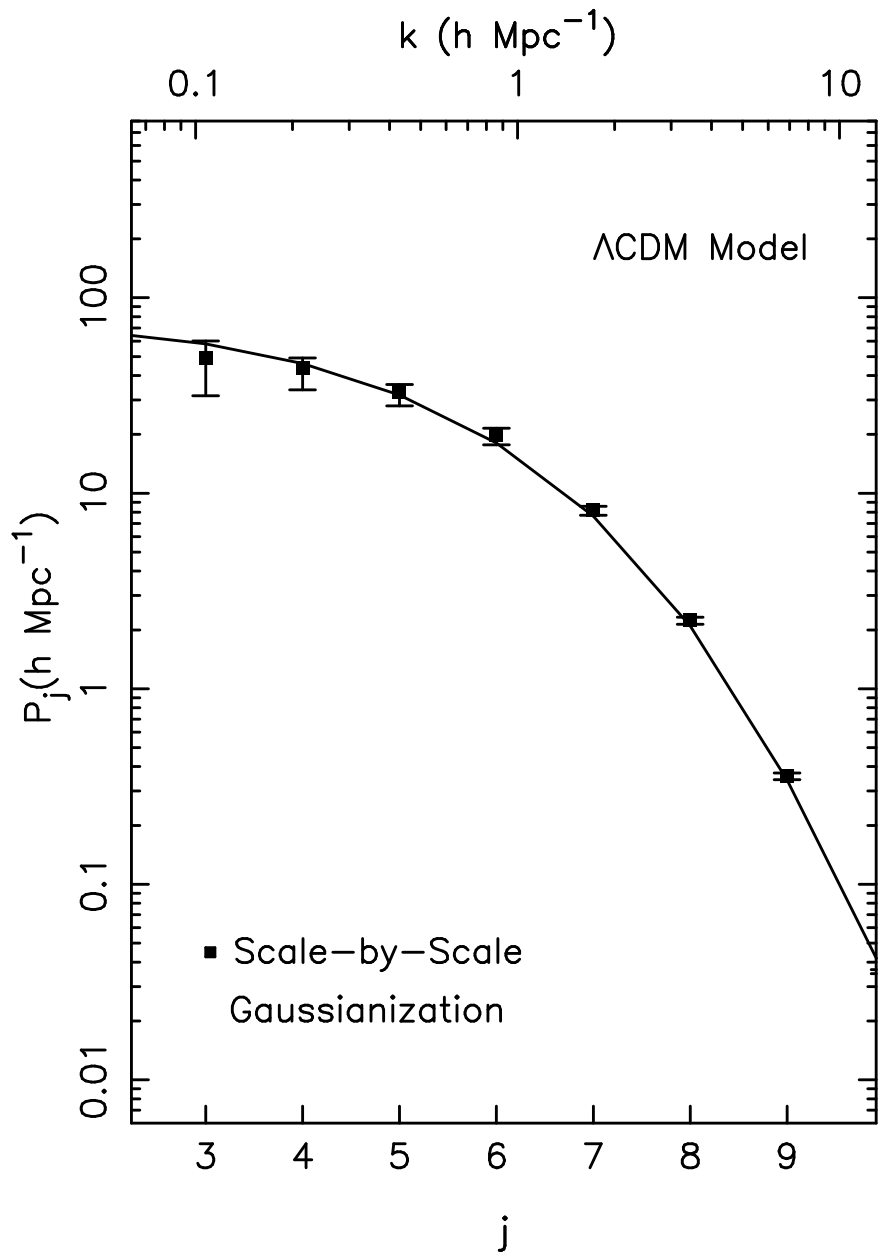
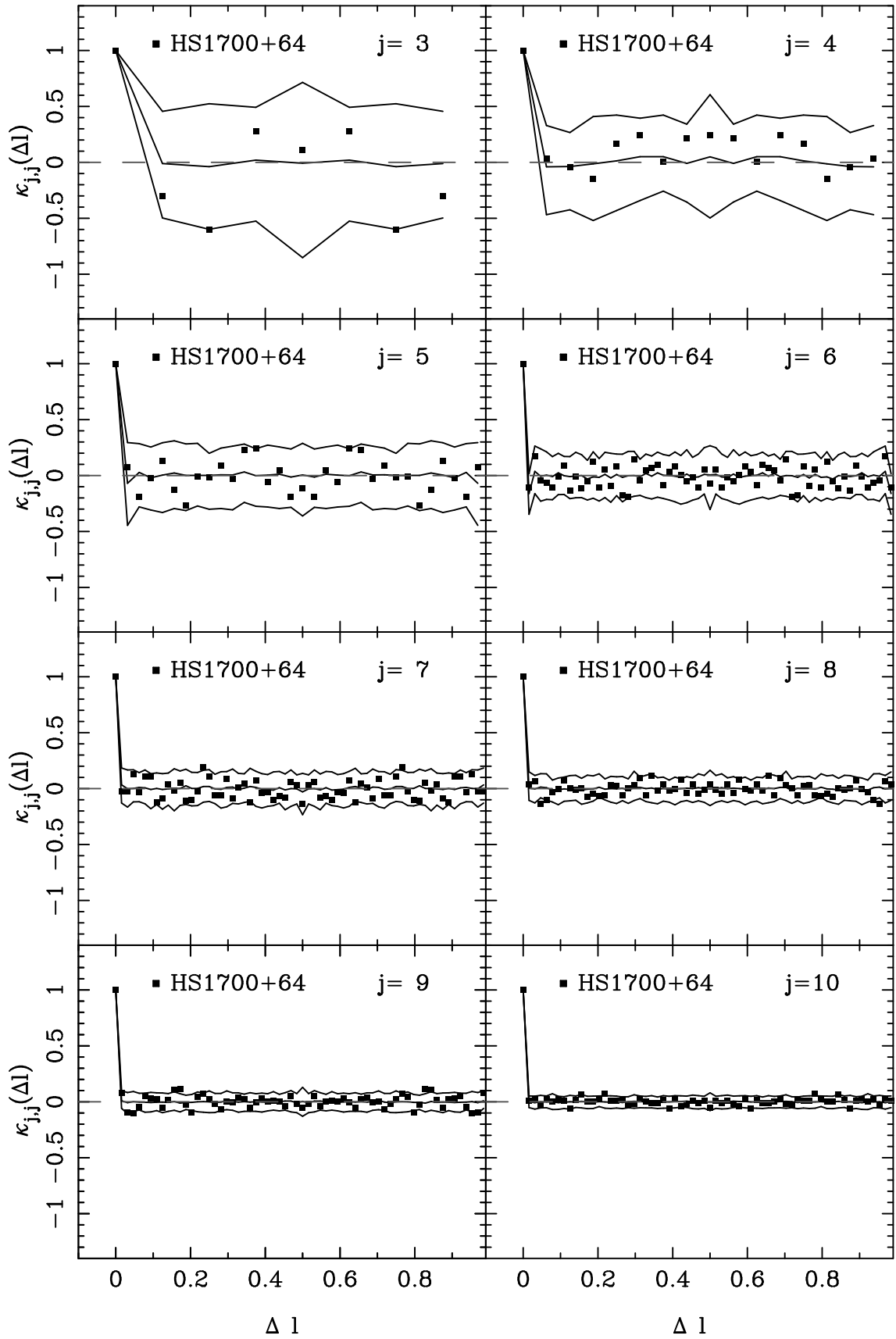
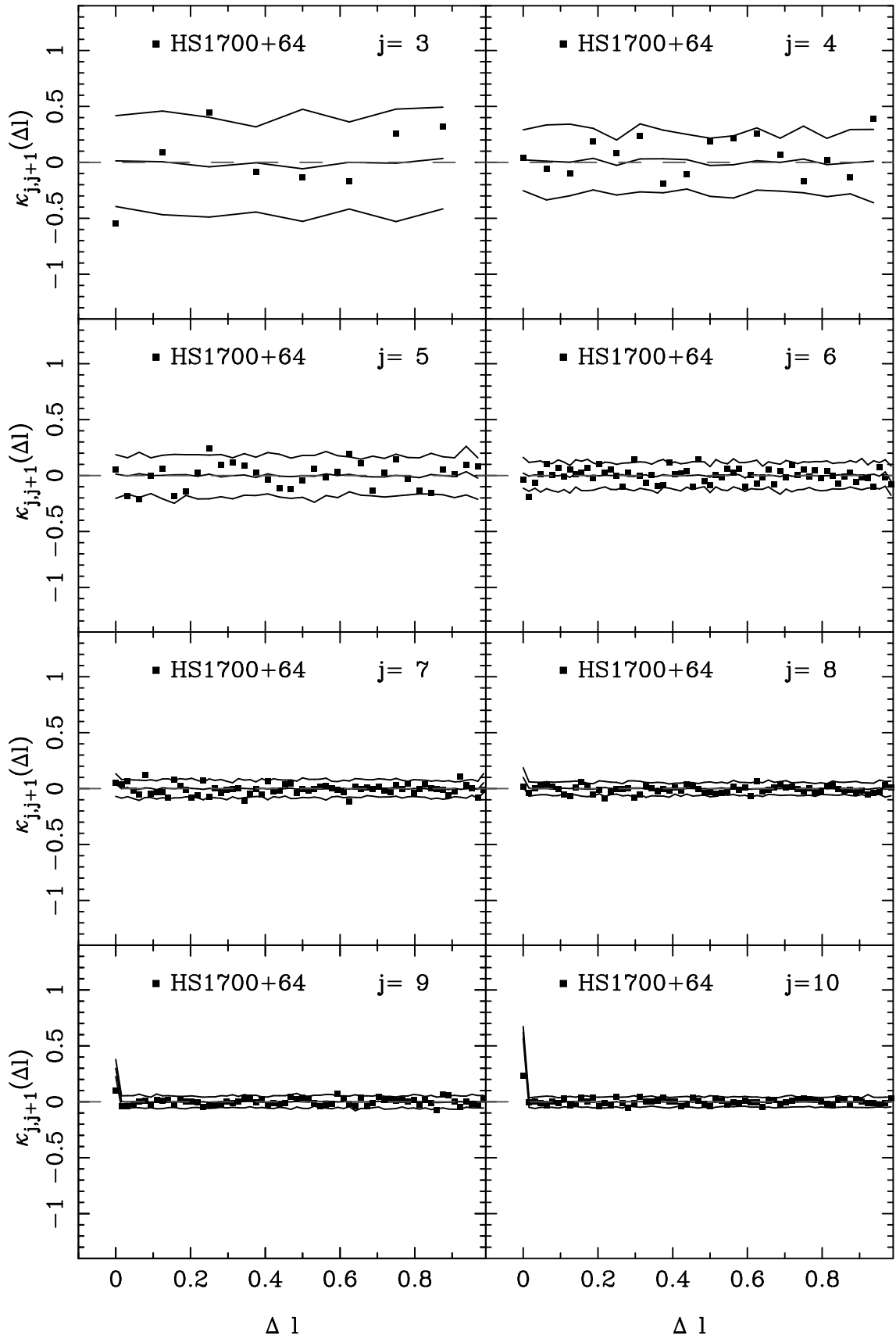
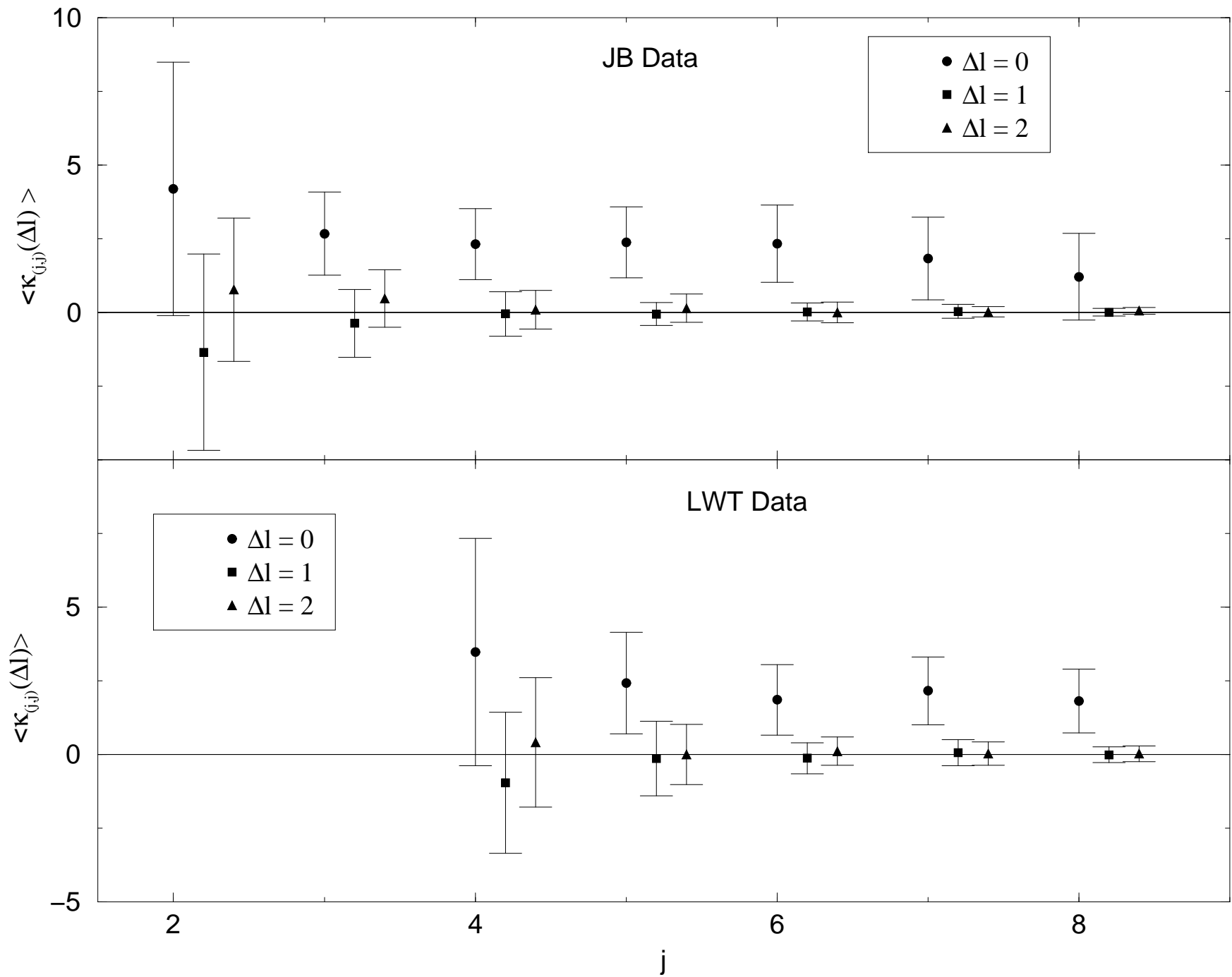
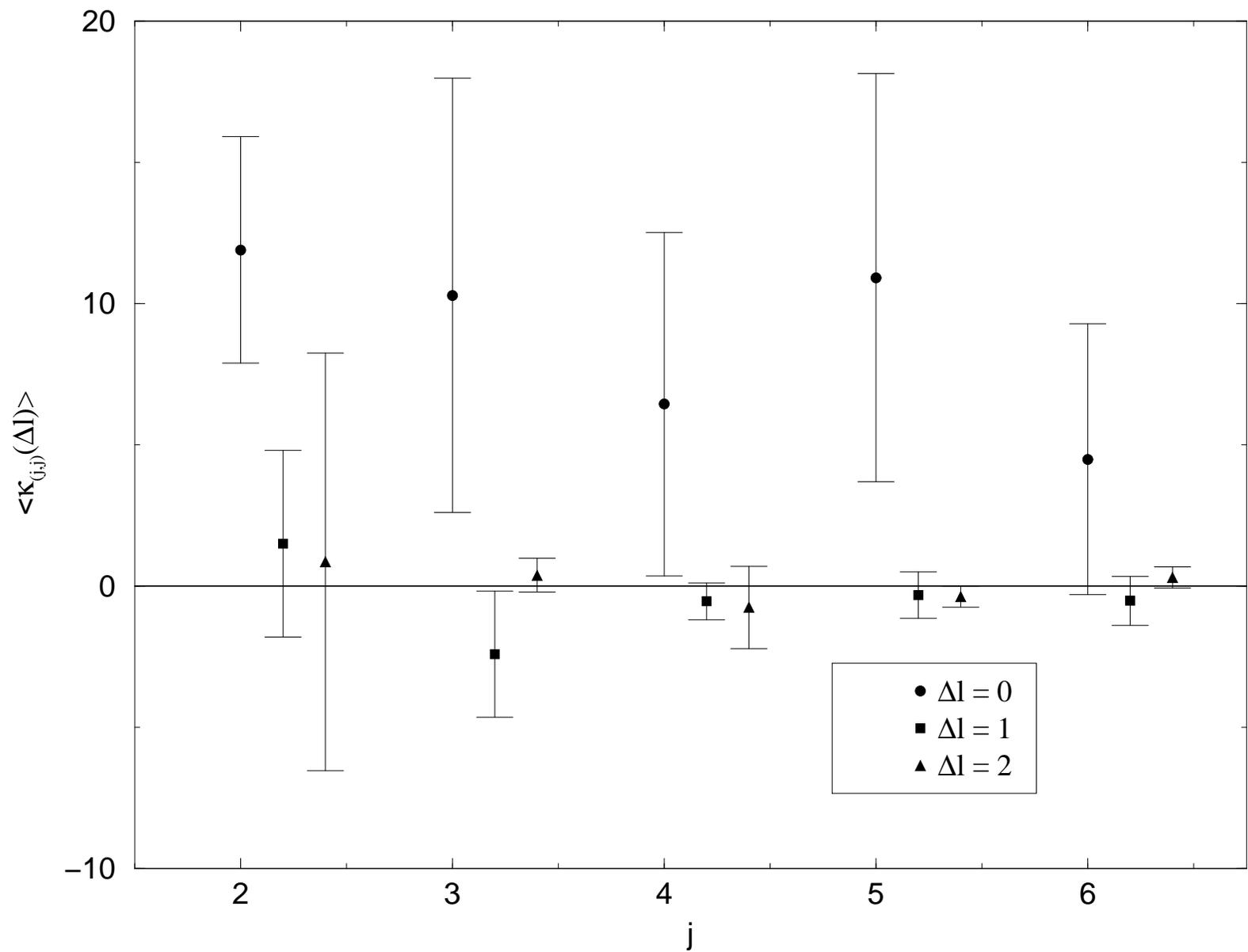
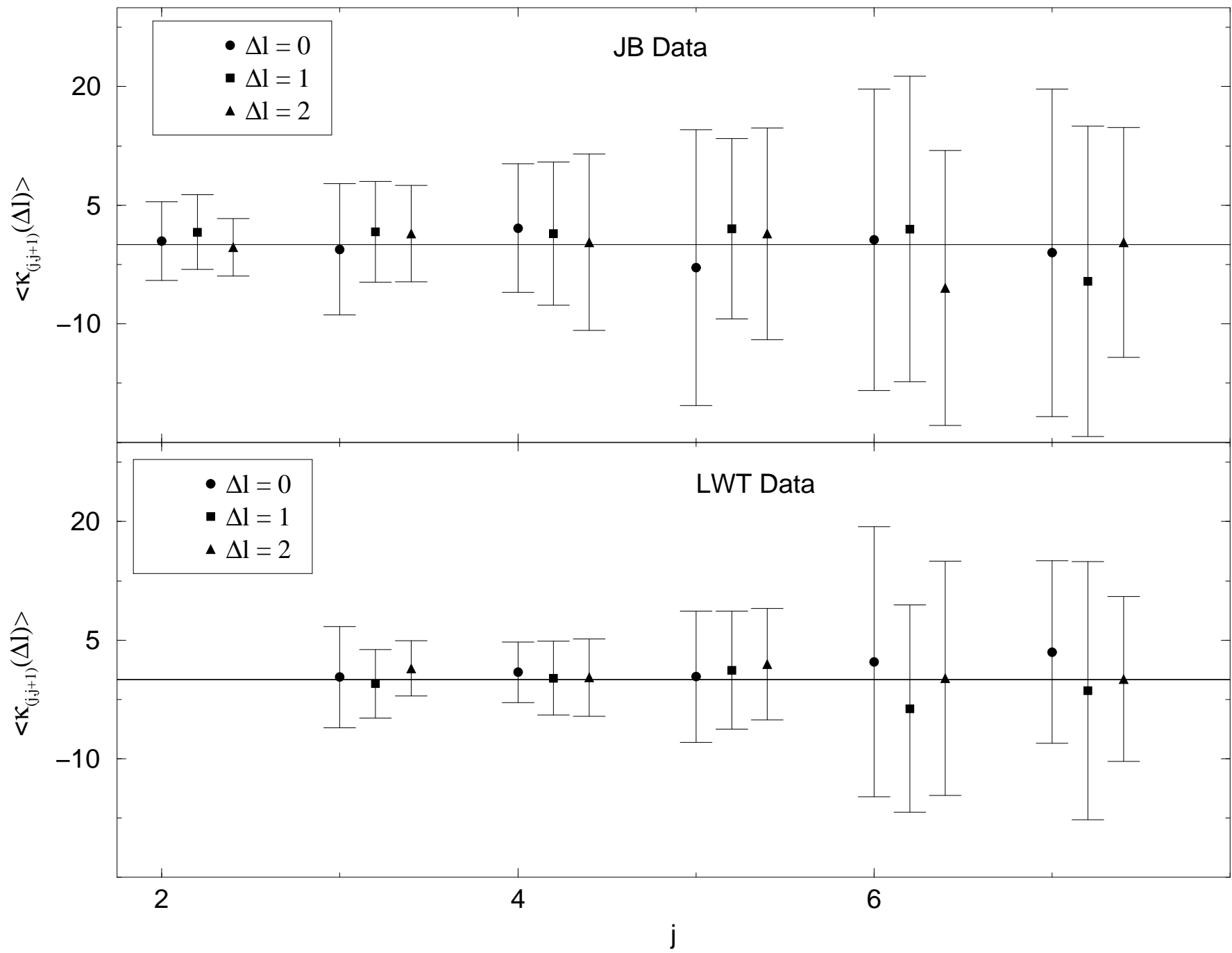
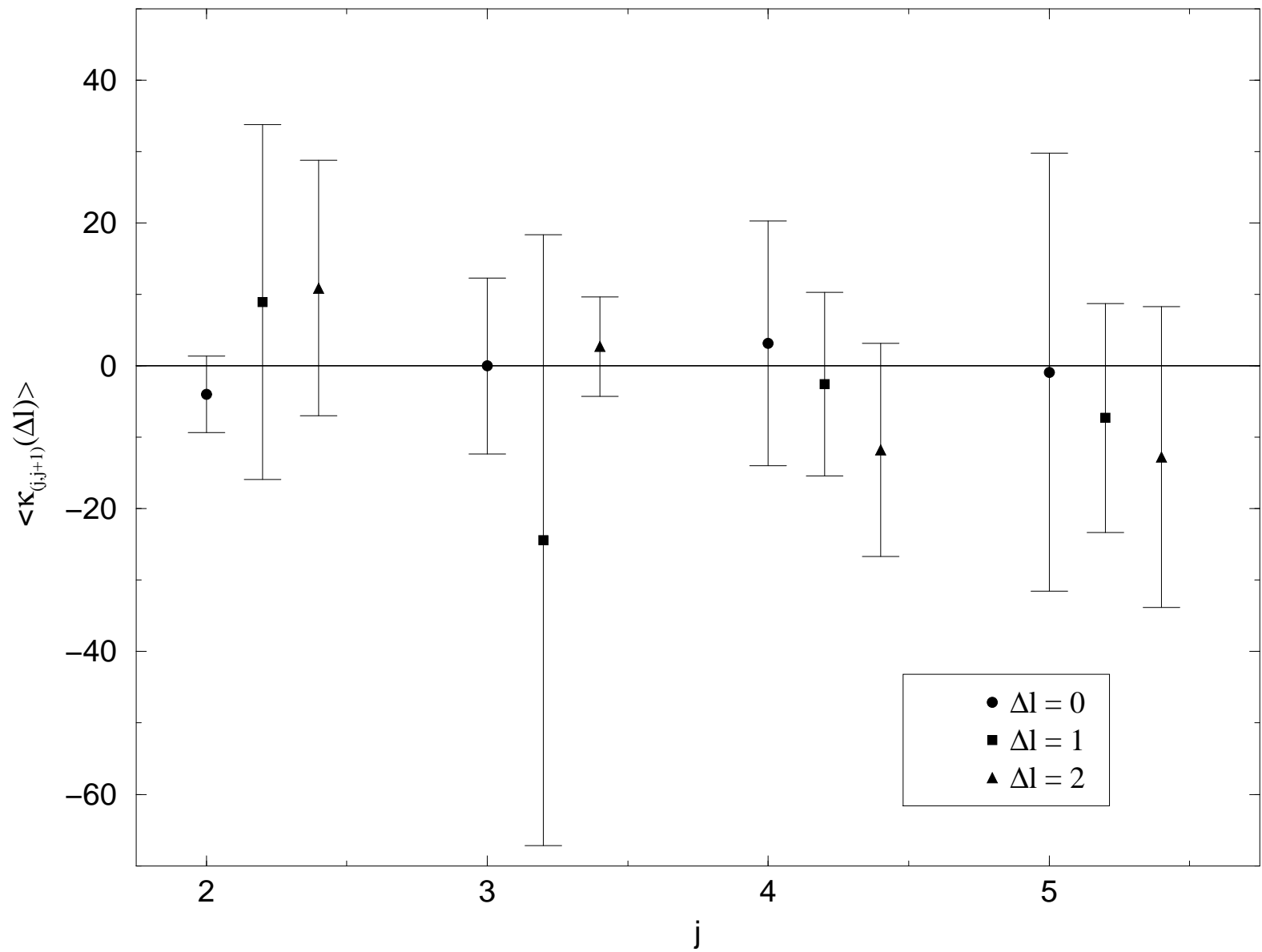
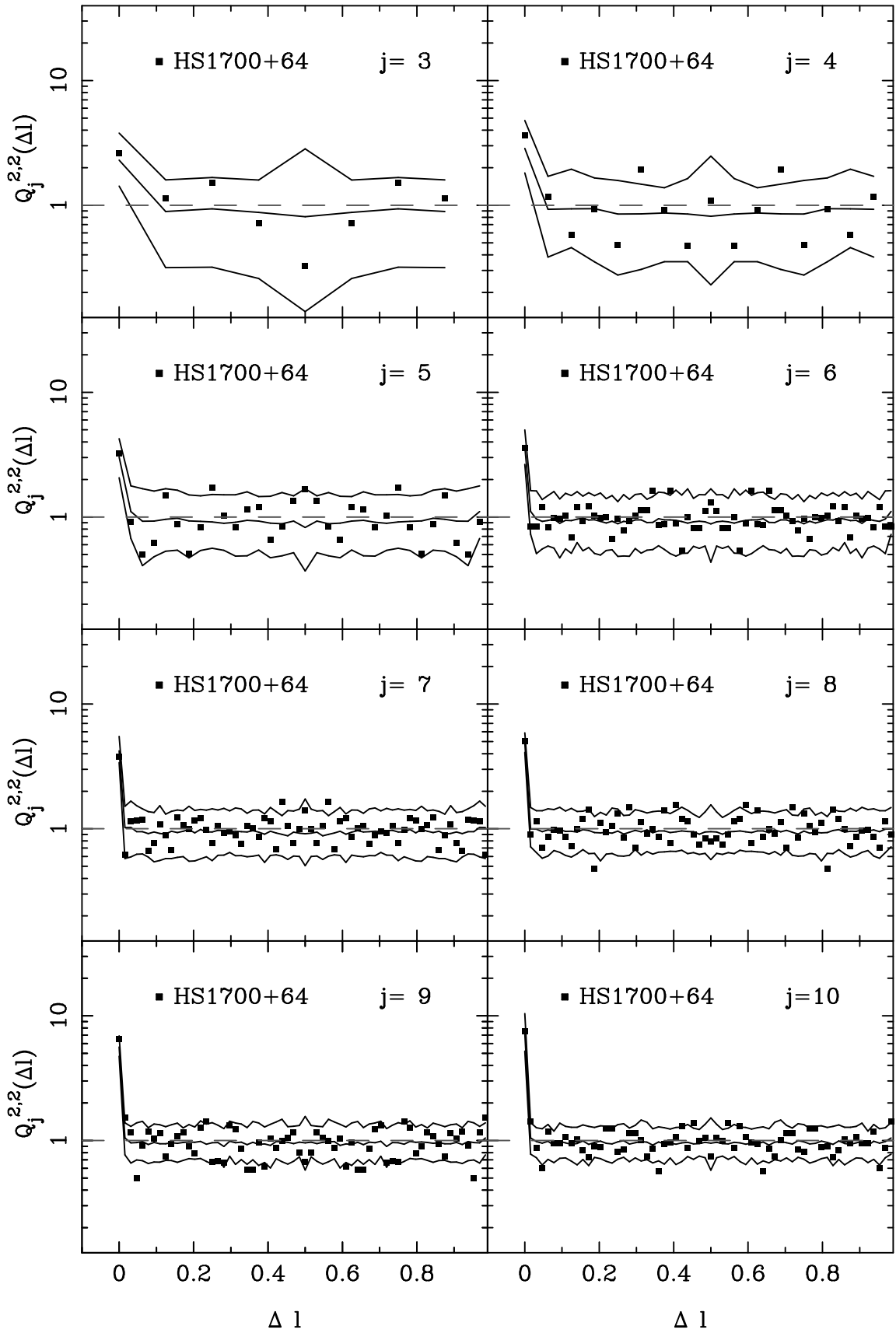
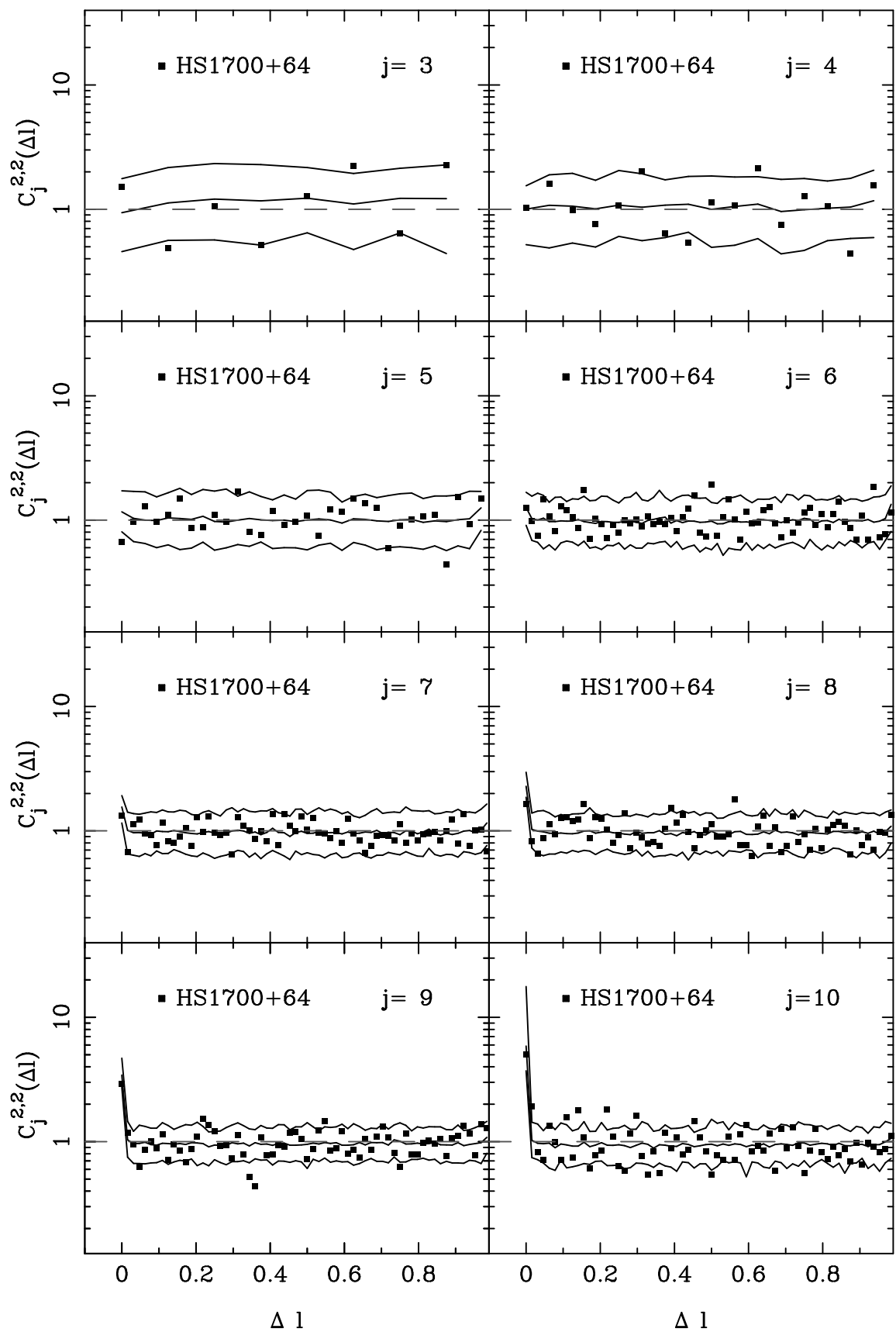
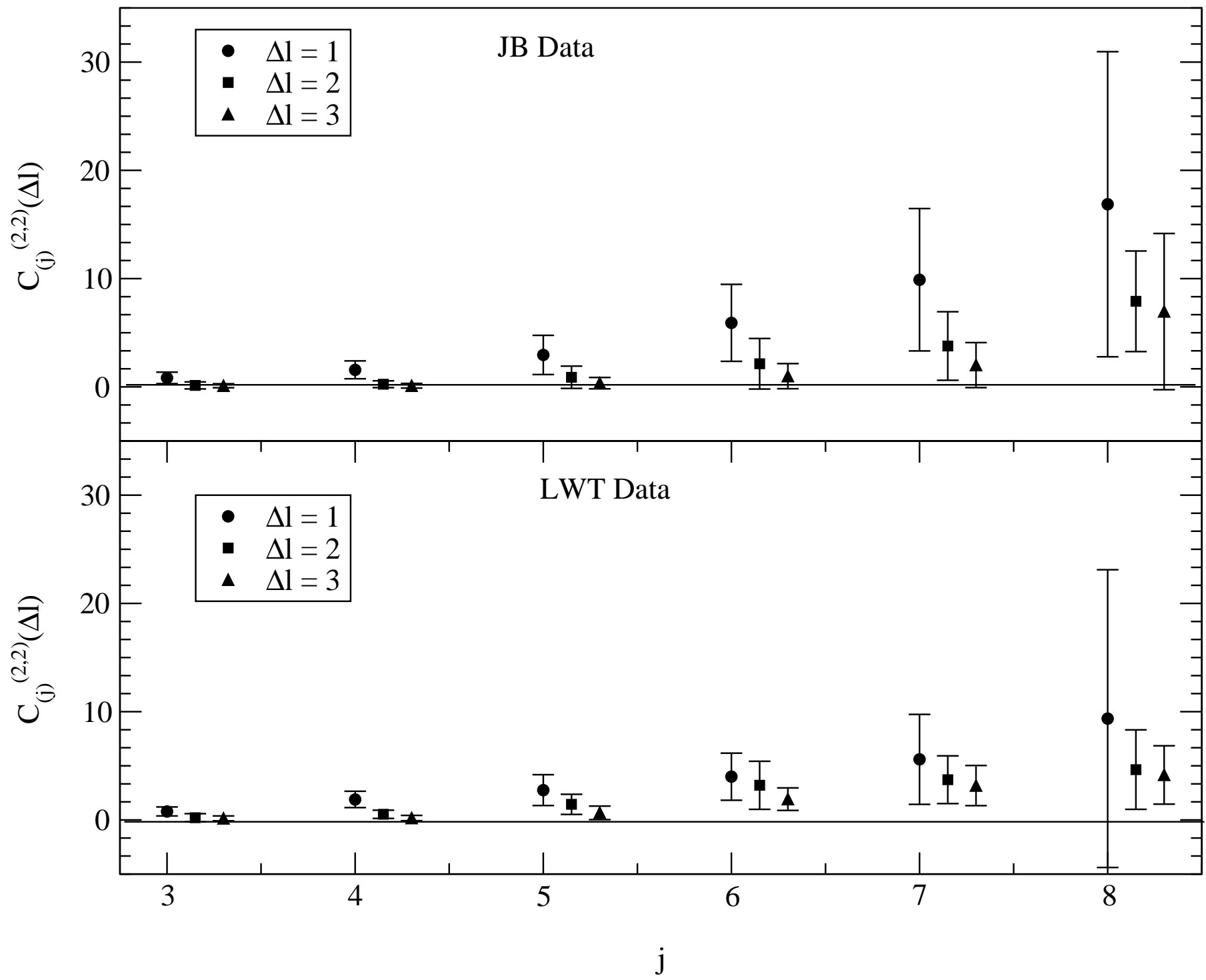
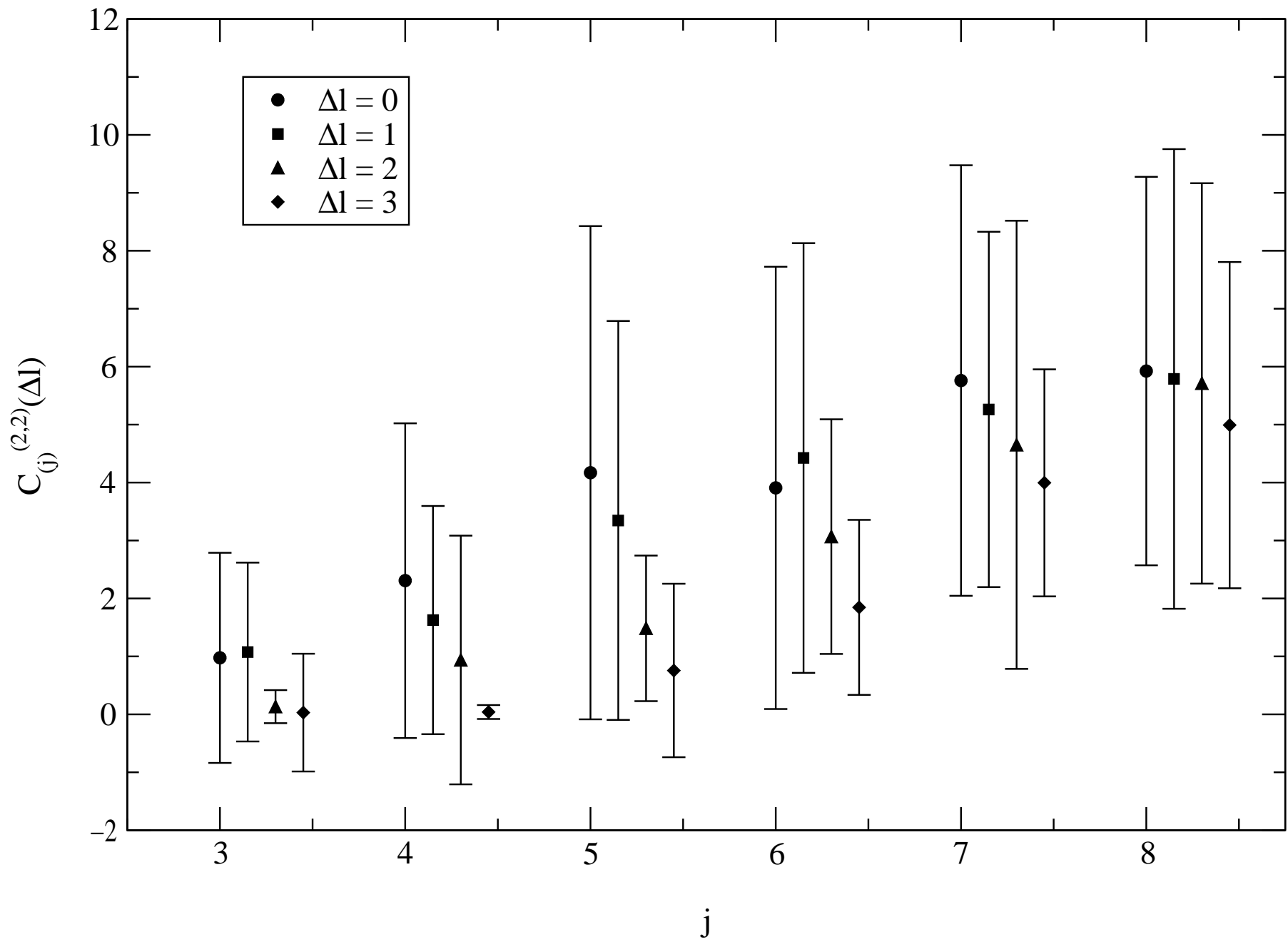
\end{document}